\documentclass[twocolumn,twocolappendix]{aastex701}

\usepackage{amsmath,amssymb,bm}
\usepackage{mathtools}
\usepackage{microtype}
\usepackage{booktabs}
\usepackage{enumitem}

\newcommand{\dd}{\mathrm{d}}
\newcommand{\vct}[1]{\bm{#1}}
\newcommand{\mtx}[1]{\bm{#1}}
\newcommand{\I}{\mtx{I}}
\newcommand{\Bcross}{\mtx{B}_{\times}}
\newcommand{\OmL}{\Omega_{\rm L}}
\newcommand{\ts}{t_{\rm s}}
\newcommand{\pg}{\vct{p}_{\rm g}}

\newcommand{\vg}{\vct{v}_{\rm g}}
\newcommand{\vd}{\vct{v}_{\rm d}}
\newcommand{\qrel}{\vct{q}}
\newcommand{\Wrel}{\vct{W}}

\shorttitle{Nonlinear GIRK for Multifluid Charged Dust}
\shortauthors{Chen, Li, He, Wibking, Krumholz, \& Bai}

\begin{document}

\title{A Nonlinear General Implicit Runge--Kutta Integrator for Multifluid Charged Dust and Gas}

\correspondingauthor{Hui Li \& Chong-Chong He}

\author[0009-0007-1246-0222]{Mengxuan Chen}
\affiliation{Department of Astronomy, Tsinghua University, Beijing, 100084, China}
\email{mx-chen24@mails.tsinghua.edu.cn}

\author[0000-0002-1253-2763]{Hui Li}
\affiliation{Department of Astronomy, Tsinghua University, Beijing, 100084, China}
\email[show]{hliastro@tsinghua.edu.cn}

\author[0000-0002-2332-8178]{Chong-Chong He}
\affiliation{Research School of Astronomy and Astrophysics, Australian National University, Canberra, ACT 2611, Australia}
\email[show]{Chongchong.He@anu.edu.au}

\author[0000-0003-3175-2291]{Benjamin D.~Wibking}
\affiliation{Department of Physics and Astronomy, Michigan State University, East Lansing, MI 48824, USA}
\email{wibkingb@msu.edu}

\author[0000-0003-3893-854X]{Mark R.~Krumholz}
\affiliation{Research School of Astronomy and Astrophysics, Australian National University, Canberra, ACT 2611, Australia}
\email{mark.krumholz@anu.edu.au}

\author[0000-0001-6906-9549]{Xue-Ning Bai}
\affiliation{Department of Astronomy, Tsinghua University, Beijing, 100084, China}
\affiliation{Institute for Advanced Study, Tsinghua University, Beijing, 100084, China}
\email{xbai@tsinghua.edu.cn}

\begin{abstract}
Charged dust in magnetized gas is subject to aerodynamic drag and the Lorentz force,
both of which can be difficult to model numerically in a fluid simulation where their
characteristic timescales are shorter than the timestep set by the Courant condition.
We present a two-stage general implicit Runge--Kutta (GIRK) source integrator for
multifluid charged dust coupled to an ideal magnetohydrodynamic (MHD) gas, implemented
in the open source GPU-accelerated \textsc{quokka} code. Drag and Lorentz
forces are advanced together in one implicit source update, avoiding errors in the
terminal drift that can arise when the two forces are advanced in separate substeps. The
timestep is compared with a drag--Lorentz timescale to select between resolved and stiff
branches. The resolved branch uses Gauss--Legendre coefficients, while the stiff branch
uses a coefficient set designed to recover the terminal drift set by the drag--Lorentz
balance. We also allow the dust stopping time and charge-to-mass ratio to depend on the
evolving state, evaluating them separately at the two implicit stage states and solving
the resulting nonlinear stage equations by Picard iteration. The source update conserves
total gas--dust momentum, and its computational cost scales linearly with the number of
dust species. Tests reproduce the expected drag damping and terminal drift, confirm the
predicted gyromotion amplitude and phase errors, demonstrate accurate treatment of
state-dependent source coefficients, and recover the expected wave, shock, and
instability dynamics in dusty MHD flows.
\end{abstract}


\section{Introduction}\label{sec:intro}

Dust, despite its relatively low abundance, plays a key role in physical and chemical processes
across a wide range of astrophysical environments. In the diffuse interstellar medium,
star-forming molecular clouds, and protoplanetary disks, dust controls extinction
through absorption and scattering, produces infrared and submillimeter emission,
drives photoelectric heating, provides surfaces for chemistry, depletes refractory
elements from the gas phase, and supplies the opacity that couples radiation to gas
\citep[e.g.,][]{Draine2004,Draine2011}. The grain size distribution, shaped by dust growth
and destruction, influences both the radiative properties and dynamics of dusty systems
\citep[e.g.,][]{Birnstiel2024,Dubois2024,Parente2026}. In particular, the dynamical
coupling between dust and gas depends strongly on grain size because the aerodynamic
stopping time varies with grain radius and local gas conditions. In protoplanetary
disks, radial drift redistributes solids \citep{Weidenschilling1977}, while streaming
instabilities \citep{YoudinGoodman2005,Johansen2007} and turbulent concentration
\citep{Cuzzi2008} can enhance local solid densities and promote planetesimal formation.
In turbulent molecular clouds, grains of different sizes respond differently to gas
motions, producing spatial variations in the dust-to-gas mass ratio and grain size
distribution \citep[e.g.,][]{HopkinsLee2016,LeeHopkinsSquire2017,Commercon2023} that can
in turn alter dust extinction curves.

Dust grains in many astrophysical environments become electrically charged through
collisions with electrons and ions \citep{DraineSutin1987} and through photoelectric
emission induced by ambient radiation fields \citep{WeingartnerDraine2001}. In a
magnetized gas, the Lorentz force changes the spatial distribution of charged dust in
magnetized turbulence \citep{LeeHopkinsSquire2017} and affects the distribution of
dust--gas drift velocities \citep{Moseley2025}. Charged grains can also modify the
structure of magnetized shocks through their coupling to the magnetic field and their
collisional interaction with the neutral gas \citep{Draine1980,Wardle1998}. The
gyromotion of charged grains allows resonant interactions with magnetohydrodynamic (MHD)
waves, which can lead to wave growth or damping \citep{HopkinsSquire2018}.
Such interactions may affect cosmic ray transport by changing the amplitudes of
Alfv\'en waves that scatter cosmic rays \citep{Squire2021}. Magnetic
effects on the dust become especially important when the dust gyrofrequency is
comparable to, or larger than, the inverse stopping time. In this regime, drag damping
and magnetic rotation jointly determine the dust--gas relative motion. 

However, these processes can be difficult to model numerically because both drag and
gyromotion can be stiff. The stopping
time \(\ts\) and the inverse dust gyrofrequency \(|\OmL|^{-1}\) can be much shorter than
the fluid timestep set by the Courant condition, making explicit integration inefficient
or unstable. A second difficulty is specific to stiff charged dust: the Lorentz force
rotates the dust--gas relative velocity, so the terminal drift is generally not parallel
to the external forcing that drives the dust relative to the gas. It is instead
determined by a balance between this differential forcing, aerodynamic drag, and the
Lorentz force, which we refer to as the drag--Lorentz balance. A method that advances
drag and Lorentz forces in separate source substeps can therefore approach the wrong
terminal drift even if each substep is stable.

There are several numerical methods available that adopt different approaches to
these challenges. For neutral dust dynamics, commonly used methods include one-fluid,
superparticle, and multifluid approaches. Full one-fluid formulations describe the
gas--dust mixture as a single fluid moving with the barycentric velocity, while
evolving the dust fractions and dust--gas differential velocities
\citep{LaibePrice2014a,LaibePrice2014b}, with extensions to multiple grain species
\citep{LaibePrice2014c,Hutchison2026}. In the tightly coupled limit, the terminal
velocity approximation replaces the differential velocity evolution with a local
force balance, allowing efficient integration of the dynamics of small grains
\citep{PriceLaibe2015,Tricco2017,Lebreuilly2019}. However, one-fluid methods cannot
represent crossing dust streams because they assign a single velocity to each dust
species at a given position.

A second class treats dust as Lagrangian superparticles coupled to the gas. This
approach naturally represents broad grain size distributions and weakly coupled or
crossing dust trajectories \citep[e.g.,][]{HopkinsLee2016,McKinnon2018}. In tightly coupled
regimes, dust superparticles can become artificially trapped if the gas does not resolve
the stopping length \(c_s\ts\) or is more coarsely resolved than the dust, leading to
excessive dust concentrations \citep{LaibePrice2012a,Tricco2017,Commercon2023}.

A third class is the multifluid approach, in which each grain size bin is represented by
a separate pressureless dust fluid coupled to the gas by drag source terms. Such methods
have been formulated in smoothed particle hydrodynamics
\citep{LaibePrice2012a,LaibePrice2012b,Mentiplay2020} and on Eulerian meshes
\citep{Miniati2010,BenitezLlambay2019}. This approach is
well suited to simulations in which a finite number of grain size bins adequately
describes the dust population. Recent Eulerian implementations include implicit and
semi-implicit drag solvers that remain stable when the stopping time is shorter than the
fluid timestep \citep{HuangBai2022,Krapp2024,TedeschiPrades2025}.
Methods based on the matrix exponential provide another approach to the stiff drag
problem \citep{Sewanou2025}. To reduce spurious dust-to-gas ratio variations arising
from numerical transport in the strong-coupling limit, \citet{Verrier2025,Iwasaki2026}
construct dust mass fluxes consistent with gas transport, even when the stopping length
is unresolved.

Numerical treatments of charged dust are less extensive than those developed for neutral
dust. In applications to the interstellar medium and star formation, charged grains have
often been treated as particles or superparticles subject to drag and Lorentz forces
\citep{LeeHopkinsSquire2017,Moseley2025}.
For example, \citet{Moseley2023} implemented a magnetohydrodynamic particle-in-cell
(MHD-PIC) treatment in \textsc{ramses} in which dust superparticles experience both
aerodynamic drag and Lorentz forces, with the two source terms advanced by separate
implicit or semi-implicit substeps. \citet{BeitiaAntero2021} used a semi-implicit particle
update that treats aerodynamic drag, Coulomb drag, and the Lorentz force together.
Charged grains have also been modeled as separate fluids in multifluid calculations
of shocks in weakly ionized gas \citep{CiolekRoberge2002,Chapman2006,Guillet2007}.

A further complication arises when the dust stopping time or charge-to-mass ratio varies
during a source step. The stopping time can depend on the gas density, sound speed, and
dust--gas drift speed, while the grain charge can depend on the local plasma and
thermodynamic state through the adopted charging model. The source equations are then
nonlinear, so the coefficients entering the source terms must be evaluated consistently
with the evolving state during the implicit update.

Motivated by these numerical issues, we develop a two-stage general implicit
Runge--Kutta (GIRK) source integrator for multifluid charged dust in ideal
MHD simulations and implement it in \textsc{quokka}, a
GPU-accelerated radiation-MHD code with adaptive mesh refinement
\citep{WibkingKrumholz2022,He2024a,He2024b,Kriel2026}. The method builds on the
neutral multifluid GIRK drag integrator of \citet{TedeschiPrades2025}. For charged
dust, aerodynamic drag and the Lorentz force are advanced together in a single
implicit solve, avoiding an explicit timestep restriction from drag relaxation or
gyromotion. For constant source coefficients, the stiff branch approaches the
terminal drift determined by the drag--Lorentz balance when the gas and dust
experience different external forces. We also extend the method to
state-dependent stopping times and charge-to-mass
ratios. The source coefficients are evaluated separately at the two stage
states, and the resulting nonlinear equations are solved by Picard iteration. The
computational cost remains linear in the number of dust species, and the update
conserves total gas--dust momentum.

This paper is organized as follows: Section~\ref{sec:method} presents the governing
equations, describes the source integrator allowing for state-dependent source
coefficients, and analyzes its gyromotion errors and terminal drift behavior.
Section~\ref{sec:tests} tests the source update and presents applications to dusty MHD
waves, shocks, and a magnetized resonant drag instability (RDI).
Section~\ref{sec:summary_discussion} discusses the limitations of the present model and
possible extensions and summarizes the main results.

\section{Methods}\label{sec:method}

\subsection{Governing equations for the dusty ideal MHD system}
\label{subsec:governing}

We consider a magnetized gas described by the ideal MHD equations, coupled to
\(N_{\rm d}\) pressureless dust species. The dust species exchange momentum with the gas
through aerodynamic drag, and charged grains additionally experience the Lorentz force.
We use Heaviside--Lorentz units. The governing equations are
\begin{align}
&\frac{\partial \rho_{\rm g}}{\partial t}
  + \nabla\cdot(\rho_{\rm g}\vg)
  =0, \\
\frac{\partial (\rho_{\rm g}\vg)}{\partial t}
  + \nabla\cdot&\left[
      \rho_{\rm g}\vg\vg
      + \left(P_{\rm g}+\frac{|\vct{B}|^2}{2}\right)\I
      - \vct{B}\vct{B}
    \right]
  = \notag\\
  \sum_{i=1}^{N_{\rm d}}
  \rho_{{\rm d},i}
  \frac{\vct{v}_{{\rm d},i}-\vg}{t_{{\rm s},i}}
  &- \sum_{i=1}^{N_{\rm d}}
  \rho_{{\rm d},i}\Omega_{{\rm L},i}
  \left(\vct{v}_{{\rm d},i}-\vg\right)\times\hat{\vct{b}}
  + \rho_{\rm g}\vct{a}_{\rm ext,g}, \\
\frac{\partial E_{\rm g}}{\partial t}
  + \nabla\cdot&\left[
      \left(E_{\rm g}+P_{\rm g}+\frac{|\vct{B}|^2}{2}\right)\vg
      -(\vg\cdot\vct{B})\vct{B}
    \right]
  = \notag\\
  \sum_{i=1}^{N_{\rm d}}
  \rho_{{\rm d},i}
  \frac{\vct{v}_{{\rm d},i}-\vg}{t_{{\rm s},i}}&\cdot\vg
  -\sum_{i=1}^{N_{\rm d}}
  \rho_{{\rm d},i}\Omega_{{\rm L},i}
  \left[
    \left(\vct{v}_{{\rm d},i}-\vg\right)\times\hat{\vct{b}}
  \right]\cdot\vg
  \notag\\
    + \rho_{\rm g}\vct{a}_{\rm ext,g}\cdot\vg
  &+ \omega_{\rm drag}
  \sum_{i=1}^{N_{\rm d}}
  \rho_{{\rm d},i}
  \frac{\left|\vct{v}_{{\rm d},i}-\vg\right|^2}{t_{{\rm s},i}},
  \label{eq:gas_energy_continuum}\\
&\frac{\partial \vct{B}}{\partial t}
  - \nabla\times(\vg\times\vct{B})
  =0, \\
&\frac{\partial \rho_{{\rm d},i}}{\partial t}
  + \nabla\cdot(\rho_{{\rm d},i}\vct{v}_{{\rm d},i})
  =0, \\
\frac{\partial(\rho_{{\rm d},i}\vct{v}_{{\rm d},i})}{\partial t}
  &+ \nabla\cdot
  \left(
    \rho_{{\rm d},i}
    \vct{v}_{{\rm d},i}\vct{v}_{{\rm d},i}
  \right)
  = \rho_{{\rm d},i}
  \frac{\vg-\vct{v}_{{\rm d},i}}{t_{{\rm s},i}}
  \notag\\
  +\rho_{{\rm d},i}\Omega_{{\rm L},i}
  &\left(\vct{v}_{{\rm d},i}-\vg\right)\times\hat{\vct{b}}
  + \rho_{{\rm d},i}\vct{a}_{{\rm ext,d},i}.
\end{align}
Here \(E_{\rm g}=\rho_{\rm g}e_{\rm g}+\rho_{\rm g}|\vg|^2/2+|\vct B|^2/2\) is the
gas total energy density, where \(e_{\rm g}\) is the gas specific internal energy. The
magnetic field satisfies \(\nabla\cdot\vct B=0\). The stopping time of dust species
\(i\) is \(t_{{\rm s},i}\). Let \(q_{{\rm gr},i}\) and \(m_{{\rm gr},i}\) denote the
signed charge and mass of a grain of species \(i\), respectively. The ideal MHD electric
field is \(\vct{E}=-(\vg/c)\times\vct{B}\), so the Lorentz acceleration entering the
second source term of the dust momentum equation above is
\([q_{{\rm gr},i}/(m_{{\rm gr},i} c)]
(\vct{v}_{{\rm d},i}-\vg)\times\vct{B}\). For \(|\vct{B}|>0\), we define
\begin{equation}
  \hat{\vct b}=\frac{\vct{B}}{|\vct{B}|},
  \qquad
  \Omega_{{\rm L},i}
  =
  \frac{q_{{\rm gr},i}|\vct{B}|}{m_{{\rm gr},i} c}.
  \label{eq:gyrofrequency}
\end{equation}
Thus \(\Omega_{{\rm L},i}\) is the signed grain gyrofrequency: its magnitude sets the
rate of gyromotion, while its sign determines the direction of rotation. For
\(\vct{B}=0\), the Lorentz term is zero and \(\hat{\vct b}\) is not required. The equal
and opposite Lorentz term appears in the gas momentum equation as the dust
back-reaction.

The first three source terms in Equation~\eqref{eq:gas_energy_continuum} describe
mechanical work by aerodynamic drag, the Lorentz back-reaction from charged dust, and
the external gas acceleration. The factor \(\omega_{\rm drag}\) specifies the fraction
of the aerodynamic drag dissipation that is deposited as gas internal energy in this
model. The discrete gas energy update is described in Section~\ref{subsec:energy}. In a
model that also solves the thermal equilibrium equations coupling dust, gas, and
radiation, the remaining fraction \(1-\omega_{\rm drag}\) of the drag dissipation would
be included as a heating term in these equations, so that the drag dissipation remains
energy conserving.

\subsection{Units}
\label{subsec:dimensionless_units}

The numerical tests in Section~\ref{sec:tests} use dimensionless variables defined by
reference length, density, and time scales \(L_0\), \(\rho_0\), and \(\tau_0\). The
corresponding velocity and magnetic field scales are
\begin{equation}
  v_0=\frac{L_0}{\tau_0},
  \qquad
  B_0=\sqrt{\rho_0}\,v_0.
  \label{eq:test_units}
\end{equation}

Following \citet{Moseley2023}, we define the signed dimensionless
charge-to-mass ratio
\begin{equation}
  \xi_i
  =
  \frac{q_{{\rm gr},i}L_0\sqrt{\rho_0}}{m_{{\rm gr},i} c}.
  \label{eq:xi}
\end{equation}
Defining \(\widetilde{\vct B}=\vct B/B_0\) and
\(\widetilde{\Omega}_{{\rm L},i}=\tau_0\Omega_{{\rm L},i}\), the signed dimensionless
angular gyrofrequency satisfies
\(\widetilde{\Omega}_{{\rm L},i}=\xi_i|\widetilde{\vct B}|\). We suppress tildes on
dimensionless quantities in Section~\ref{sec:tests}.

\subsection{Strang splitting and source equations}
\label{subsec:strang_source_girk}

We advance the coupled gas--dust system using Strang splitting \citep{Strang1968}
\begin{equation}
  \mathcal{U}^{n+1}
  =
  \mathcal{C}_{\Delta t_{\mathcal C}}
  \mathcal{H}_{\Delta t}
  \mathcal{C}_{\Delta t_{\mathcal C}}
  \mathcal{U}^{n},
  \qquad
  \Delta t_{\mathcal C}\equiv\frac{\Delta t}{2}.
  \label{eq:strang}
\end{equation}
Here \(\Delta t\) is the full Strang timestep and \(\Delta t_{\mathcal C}\) is the
duration of each \(\mathcal C\) update. The operator \(\mathcal C\) denotes the combined
source update for aerodynamic drag and the Lorentz force. This drag--Lorentz update also
includes the associated gas energy update described in Section~\ref{subsec:energy}. The
operator \(\mathcal H\) advances the gas and magnetic field under ideal MHD, together
with pressureless dust transport and external accelerations. This splitting is the
charged dust counterpart of the \(\mathcal D\mathcal H\mathcal D\) splitting of
\citet{TedeschiPrades2025}. We adopt this form to place the source half-steps around the
existing \textsc{quokka} MHD and dust transport update.

The transport timestep is limited by the Courant--Friedrichs--Lewy (CFL) condition,
\begin{equation}
  \Delta t_{\rm CFL}
  =
  C_{\rm CFL}
  \min_{\rm cells}
  \frac{\Delta x_{\min}}
  {c_{\rm sig}
  +\max\left\{
    |\vg|,
    \max_i|\vct{v}_{{\rm d},i}|
  \right\}},
  \label{eq:dust_cfl}
\end{equation}
where \(C_{\rm CFL}\) is the CFL number, \(\Delta x_{\min}\) is the minimum cell width,
and \(c_{\rm sig}\) is the gas signal speed. For hydrodynamic transport, we take
\(c_{\rm sig}\) to be the sound speed \(c_s\), while in MHD it is the maximum fast
magnetosonic speed along the coordinate directions. We integrate the drag and Lorentz
source terms implicitly, so the stopping time and gyromotion timescale impose no
additional explicit stability constraint on the timestep.

During each \(\mathcal C\) update, the gas and dust densities and the magnetic field are
held fixed, while the gas and dust momenta are updated implicitly. Because constrained
transport stores the magnetic field components at cell faces, each component entering
the implicit solve is taken as the arithmetic mean of its values on the two adjacent
faces. We write \(\pg=\rho_{\rm g}\vg\) and
\(\vct p_{{\rm d},i}=\rho_{{\rm d},i}\vct v_{{\rm d},i}\), and for each dust species
define the dust-to-gas mass ratio \(\epsilon_i=\rho_{{\rm d},i}/\rho_{\rm g}\) and the
relative momentum variable
\begin{equation}
  \qrel_i
  \equiv
  \vct{p}_{{\rm d},i}-\epsilon_i\pg
  =
  \rho_{{\rm d},i}
  \left(\vct{v}_{{\rm d},i}-\vg\right).
  \label{eq:qdef}
\end{equation}

Let \(\alpha_i\equiv t_{{\rm s},i}^{-1}\) be the drag rate. We define
\begin{equation}
\Bcross =
\begin{pmatrix}
0 & b_z & -b_y\\
-b_z & 0 & b_x\\
b_y & -b_x & 0
\end{pmatrix},
\qquad
\Bcross\vct{x}=\vct{x}\times\hat{\vct{b}},
\label{eq:Bcross}
\end{equation}
and the drag--Lorentz operator
\begin{equation}
  \mtx{T}_i
  =
  -\alpha_i\I
  +
  \Omega_{{\rm L},i}\Bcross .
  \label{eq:Ti}
\end{equation}
During a \(\mathcal C\) update, the momentum equations become
\begin{equation}
\begin{aligned}
  \frac{\dd \vct{p}_{{\rm d},i}}{\dd t}
  &=
  \mtx{T}_i\qrel_i,\\
  \frac{\dd \pg}{\dd t}
  &=
  -\sum_{i=1}^{N_{\rm d}}\mtx{T}_i\qrel_i .
\end{aligned}
\label{eq:local_source_momentum}
\end{equation}
The equal and opposite source terms conserve the total gas--dust momentum. Setting
\(\Omega_{{\rm L},i}=0\) reduces these equations to pure aerodynamic drag.

Let \(\vct U=(\pg^T,\vct p_{{\rm d},1}^T,\ldots,\vct p_{{\rm d},N_{\rm d}}^T)^T\).
When the drag rates \(\alpha_i\) and gyrofrequencies \(\Omega_{{\rm L},i}\) are constant
during a \(\mathcal C\) update, the momentum system is linear,
\begin{equation}
  \frac{\dd \vct{U}}{\dd t}
  =
  \mathcal{M}\vct{U},
  \label{eq:global_linear}
\end{equation}
with
\begin{equation}
\mathcal{M}=
\begin{pmatrix}
\sum_k\epsilon_k\mtx{T}_k & -\mtx{T}_1 & \cdots & -\mtx{T}_{N_{\rm d}}\\
-\epsilon_1\mtx{T}_1 & \mtx{T}_1 & \cdots & \mtx{0}\\
\vdots & \vdots & \ddots & \vdots\\
-\epsilon_{N_{\rm d}}\mtx{T}_{N_{\rm d}} & \mtx{0} & \cdots & \mtx{T}_{N_{\rm d}}
\end{pmatrix}.
\label{eq:Mglobal}
\end{equation}
When the coefficients in \(\mathcal M\) depend on the evolving state, the momentum
problem is nonlinear and can be written as
\begin{equation}
  \frac{\dd\vct U}{\dd t}=\vct F(\vct U,E_{\rm g}).
  \label{eq:global_nonlinear}
\end{equation}
For an adiabatic gas, these source coefficients may also depend on \(E_{\rm g}\) through
thermodynamic quantities such as the sound speed. The state-dependent stopping time and
charge-to-mass ratio are described in Section~\ref{subsec:stopping_picard}.

\subsection{Two-stage GIRK source update}
\label{subsec:girk_source_update}

With the source terms integrated implicitly, the time integrator should accurately
reproduce drag damping and gyromotion in the resolved regime and damp unresolved
relative motion in the stiff regime. In the stiff limit, the full
\(\mathcal C\mathcal H\mathcal C\) update should also recover the terminal drift set by
the drag--Lorentz balance when the gas and dust experience different external
accelerations. We therefore adopt a two-stage GIRK framework with the Runge--Kutta
(RK) coefficient set chosen separately for the resolved and stiff regimes.

We apply GIRK separately to each \(\mathcal C\) half-step in
Equation~\eqref{eq:strang}. Superscripts \(-\) and \(+\) denote
states before and after an individual source update of duration
\(\Delta t_{\mathcal C}=\Delta t/2\). The two implicit stage states are
\begin{equation}
\begin{aligned}
  \vct U_1
  &=\vct U^-+\Delta t_{\mathcal C}(\gamma_1K_1+\beta_1K_2),\\
  \vct U_2
  &=\vct U^-+\Delta t_{\mathcal C}(\beta_2K_1+\gamma_2K_2).
\end{aligned}
\label{eq:girk_stage_states}
\end{equation}
The corresponding momentum stage rates are \(K_a=\vct F(\vct U_a,E_{{\rm g},a})\),
where \(a=1,2\) labels the two GIRK stages and \(\vct F\) is defined in
Equation~\eqref{eq:global_nonlinear}. For constant source coefficients,
\(K_a=\mathcal M\vct U_a\). At each stage, the drag--Lorentz operator in
Equation~\eqref{eq:Ti} is evaluated using the corresponding \(\alpha_{a,i}\) and
\(\Omega_{{\rm L},a,i}\), so the two stages may use different drag rates and
gyrofrequencies. The momentum at the end of this source update is
\begin{equation}
  \vct U^+
  =
  \vct U^-
  +
  \Delta t_{\mathcal C}
  \left[bK_1+(1-b)K_2\right].
  \label{eq:girk_update}
\end{equation}
The RK coefficients \(\gamma_1,\gamma_2,\beta_1,\beta_2,\) and \(b\) are
specified in Section~\ref{subsubsec:girk_coefficients}.

\subsubsection{Resolved and stiff coefficient sets}
\label{subsubsec:girk_coefficients}

From the current stage values of \(\alpha_{a,i}\) and \(\Omega_{{\rm L},a,i}\), define
the drag--Lorentz timescale
\begin{equation}
  \tau_{\rm DL}
  \equiv
  \max_{a,i}
  \left(\alpha_{a,i}^2+\Omega_{{\rm L},a,i}^2\right)^{-1/2},
  \label{eq:tauDL}
\end{equation}
where the maximum is taken over both stages and dust species with positive density.
Using the full Strang timestep \(\Delta t\), we select the resolved branch when
\(\Delta t<\tau_{\rm DL}\) and the stiff branch otherwise. When all gyrofrequencies
vanish, \(\tau_{\rm DL}\) reduces to the longest stopping time.

In the resolved branch, the default coefficient set depends on whether MHD is enabled.
With MHD, we use the two-stage Gauss--Legendre coefficients
\begin{align}
  \gamma_1=\gamma_2&=\frac{1}{4},
  &
  b&=\frac{1}{2},
  \notag\\
  \beta_1&=\frac{1}{4}-\frac{\sqrt{3}}{6},
  &
  \beta_2&=\frac{1}{4}+\frac{\sqrt{3}}{6}.
  \label{eq:coeff_GL4}
\end{align}
These are the coefficients of the fourth-order two-stage Gauss--Legendre method, which
we denote GL4. In the drag-free gyromotion limit, GL4 preserves the gyromotion amplitude
exactly.

For comparison, we also use the implicit midpoint method in the same two-stage form,
\begin{equation}
  \gamma_1=\gamma_2=\beta_1=\beta_2=\frac{1}{4},
  \qquad
  b=\frac{1}{2},
  \label{eq:coeff_midpoint}
\end{equation}
which we denote Midpoint. It is second-order accurate and, like GL4, preserves the
amplitude of drag-free gyromotion.

Without MHD, we use the coefficient set of \citet{TedeschiPrades2025} as the default
in the resolved branch,
\begin{equation}
  \gamma_1=1,
  \,
  \gamma_2=0,
  \,
  \beta_1=-\frac{1}{2},
  \,
  \beta_2=\frac{2}{3},
  \,
  b=1.
  \label{eq:coeff_TP2025}
\end{equation}
We refer to this coefficient set as TP2025. It is third-order accurate for constant
source coefficients and second-order accurate for general nonlinear systems.
Unlike GL4 and Midpoint, TP2025 introduces numerical damping of drag-free gyromotion.
The tests in Sections~\ref{subsec:mixed_stopping_resolved} and~\ref{subsec:gyro_tests}
motivate these defaults: TP2025 performs better in the pure drag test, whereas GL4 is
more accurate for resolved gyromotion.

In the stiff branch, we use the same coefficient set with and without MHD, following
\citet{TedeschiPrades2025}:
\begin{equation}
  \gamma_1=1,
  \,
  \gamma_2=1,
  \,
  \beta_1=1,
  \,
  \beta_2=-1,
  \,
  b=0,
  \label{eq:coeff_large}
\end{equation}
Although first-order accurate in time, this coefficient set damps unresolved drag
relaxation and gyromotion and is designed to recover the terminal drift set by the
drag--Lorentz balance in the stiff limit, as analyzed in
Section~\ref{subsec:relative_dynamics_diagnostics}.

\subsubsection{Reduced implicit solve}
\label{subsec:reduction}

For given values of \(\alpha_{a,i}\) and \(\Omega_{{\rm L},a,i}\) at the two stages,
Equations~\eqref{eq:girk_stage_states} and \eqref{eq:local_source_momentum} form a
linear system for the two momentum stage rates \(K_1\) and \(K_2\). Writing
\(K_a=(\vct{k}_{a,{\rm g}}^T,\vct{k}_{a,{\rm d},1}^T,\ldots, \vct{k}_{a,{\rm d},N_{\rm d}}^T)^T\),
each stage rate contains three momentum components for the gas and for every dust
species. Solving the full two-stage system directly would therefore require a
\(6(N_{\rm d}+1)\times6(N_{\rm d}+1)\) linear system.

To reduce this system, we substitute the gas and dust stage momenta from
Equation~\eqref{eq:girk_stage_states} into the stage form of the relative momentum
defined in Equation~\eqref{eq:qdef},
\(\qrel_{a,i}=\vct p_{{\rm d},i,a}-\epsilon_i\vct p_{{\rm g},a}\). This expresses
\(\qrel_{a,i}\) in terms of \(\qrel_i^-\), the two dust stage rates of species \(i\),
and the two gas stage rates. Combining this expression with
\(\vct{k}_{a,{\rm d},i}=\mtx T_{a,i}\qrel_{a,i}\) from
Equation~\eqref{eq:local_source_momentum} gives a two-stage linear system for the dust
stage rates. For given gas stage rates, this system can be solved independently for each
dust species. Defining \(\mtx L_{a,i}=\Delta t_{\mathcal C}\mtx T_{a,i}\), the system is
\begin{align}
  \begin{pmatrix}
    \I-\gamma_1\mtx{L}_{1,i} & -\beta_1\mtx{L}_{1,i}\\
    -\beta_2\mtx{L}_{2,i} & \I-\gamma_2\mtx{L}_{2,i}
  \end{pmatrix}
  \begin{pmatrix}
    \vct{k}_{1,{\rm d},i}\\
    \vct{k}_{2,{\rm d},i}
  \end{pmatrix}
  =\notag\\
  \begin{pmatrix}
    \mtx{T}_{1,i}\qrel_i^-
    -\epsilon_i\mtx{L}_{1,i}
    \left(\gamma_1\vct{k}_{1,{\rm g}}
    +\beta_1\vct{k}_{2,{\rm g}}\right)\\
    \mtx{T}_{2,i}\qrel_i^-
    -\epsilon_i\mtx{L}_{2,i}
    \left(\beta_2\vct{k}_{1,{\rm g}}
    +\gamma_2\vct{k}_{2,{\rm g}}\right)
  \end{pmatrix}.
  \label{eq:dust6}
\end{align}
Solving this system gives
\begin{equation}
\begin{aligned}
  \vct{k}_{1,{\rm d},i}
  &=
  \mtx{P}_{1,i}\qrel_i^-
  + \mtx{X}_{1,i}\vct{k}_{1,{\rm g}}
  + \mtx{Y}_{1,i}\vct{k}_{2,{\rm g}},\\
  \vct{k}_{2,{\rm d},i}
  &=
  \mtx{P}_{2,i}\qrel_i^-
  + \mtx{X}_{2,i}\vct{k}_{1,{\rm g}}
  + \mtx{Y}_{2,i}\vct{k}_{2,{\rm g}}.
\end{aligned}
\label{eq:dust_affine}
\end{equation}
Momentum conservation gives \(\vct{k}_{a,{\rm g}}=-\sum_i\vct{k}_{a,{\rm d},i}\) at each
stage. Substituting Equation~\eqref{eq:dust_affine} gives the gas system
\begin{equation}
\begin{pmatrix}
\I+\sum_i\mtx{X}_{1,i} & \sum_i\mtx{Y}_{1,i}\\
\sum_i\mtx{X}_{2,i} & \I+\sum_i\mtx{Y}_{2,i}
\end{pmatrix}
\begin{pmatrix}
\vct{k}_{1,{\rm g}}\\
\vct{k}_{2,{\rm g}}
\end{pmatrix}
=
-\begin{pmatrix}
\sum_i\mtx{P}_{1,i}\qrel_i^-\\
\sum_i\mtx{P}_{2,i}\qrel_i^-
\end{pmatrix}.
\label{eq:gas6}
\end{equation}
After solving Equation~\eqref{eq:gas6}, the dust stage rates follow from
Equation~\eqref{eq:dust_affine}. Appendix~\ref{app:algebra} gives explicit formulas for
the species matrices \(\mtx P_{a,i}\), \(\mtx X_{a,i}\), and \(\mtx Y_{a,i}\), and for
solving the \(6\times6\) gas system.

Assembling the gas system in Equation~\eqref{eq:gas6} and recovering the dust stage
rates each cost \(\mathcal O(N_{\rm d})\), while solving the \(6\times6\) gas system
costs \(\mathcal O(1)\). Thus, solving for all gas and dust stage rates at fixed
source coefficients costs \(\mathcal O(N_{\rm d})\). By comparison, treating the
unreduced \(6(N_{\rm d}+1)\times6(N_{\rm d}+1)\) system as a dense linear system
would have \(\mathcal O(N_{\rm d}^3)\) computational cost.

\subsection{State-dependent source coefficients and Picard iteration}
\label{subsec:stopping_picard}

Because the densities and magnetic field do not change during a \(\mathcal C\) update,
\(\epsilon_i\) and \(\Bcross\) remain fixed. The state dependence considered here
therefore enters through the dust stopping time \(t_{{\rm s},i}\) and through the grain
charge-to-mass ratio. These respectively determine the drag rate \(\alpha_i\) and
gyrofrequency \(\Omega_{{\rm L},i}\), which enter the drag--Lorentz operator
\(\mtx T_i\) in Equation~\eqref{eq:Ti}, and hence the source matrix \(\mathcal M\) in
Equation~\eqref{eq:Mglobal}.

\subsubsection{Stopping time and charge}

For Epstein drag, the stopping time is \citep{Epstein1924,DraineSalpeter1979Drag}
\begin{equation}
  t_{{\rm s},i,0}
  =
  \frac{\sqrt{\pi\gamma}}{2\sqrt{2}}
  \frac{a_i\rho_{{\rm gr},i}}{\rho_{\rm g}c_{\rm s}},
  \label{eq:epstein_stopping}
\end{equation}
where \(a_i\) is the grain radius, \(\rho_{{\rm gr},i}\) is the grain material density,
\(c_{\rm s}\) is the gas sound speed, and \(\gamma\) is the gas adiabatic index. To
account for transonic and supersonic dust--gas drift, we use an approximation to the
Epstein--Baines drag law \citep{Baines1965,Kwok1975,DraineSalpeter1979Drag},
\begin{equation}
  t_{{\rm s},i}
  =
  t_{{\rm s},i,0}
  \left[
  1+\frac{9\pi\gamma}{128}
  \left|
  \frac{\vct{v}_{{\rm d},i}-\vg}{c_{\rm s}}
  \right|^2
  \right]^{-1/2}.
  \label{eq:epstein_baines_stopping}
\end{equation}
The correction factor approaches unity at small drift speeds and decreases the stopping time as
the drift becomes transonic or supersonic. For isothermal calculations, \(\gamma\) in
Equations~\eqref{eq:epstein_stopping} and~\eqref{eq:epstein_baines_stopping} is set to
unity.

The grain charge-to-mass ratio may also depend on local conditions such as the gas
temperature, electron and ion number densities, and radiation field
\citep{DraineSutin1987,WeingartnerDraine2001}. When the corresponding dimensionless
charge-to-mass ratio \(\xi_i\) is state dependent, we evaluate \(\xi_{a,i}\)
separately from each RK stage state, giving
\(\Omega_{{\rm L},a,i}=\xi_{a,i}|\vct B|\) in dimensionless units.

For an adiabatic gas, the gas energy is part of the nonlinear stage state because the
sound speed in Equations~\eqref{eq:epstein_stopping}
and~\eqref{eq:epstein_baines_stopping} depends on the gas thermodynamic state. If
\(\vct k_{a,{\rm g}}\) is the gas momentum rate at stage \(a\), the stage energy rate is
\begin{equation}
  K_{E,a}
  =
  \frac{\vct p_{{\rm g},a}}{\rho_{\rm g}}\cdot\vct k_{a,{\rm g}}
  +\omega_{\rm drag}
  \sum_i\frac{\alpha_{a,i}}{\rho_{{\rm d},i}}|\qrel_{a,i}|^2.
  \label{eq:stage_energy_rate}
\end{equation}
The two stage energies are constructed in the same way as the momentum stage states in
Equation~\eqref{eq:girk_stage_states},
\begin{equation}
\begin{aligned}
  E_{{\rm g},1}
  &=E_{\rm g}^-+\Delta t_{\mathcal C}
  (\gamma_1K_{E,1}+\beta_1K_{E,2}),\\
  E_{{\rm g},2}
  &=E_{\rm g}^-+\Delta t_{\mathcal C}
  (\beta_2K_{E,1}+\gamma_2K_{E,2}).
\end{aligned}
\label{eq:stage_gas_energy}
\end{equation}
These stage energies are used only to evaluate thermodynamic quantities, such as the
sound speed, needed by the state-dependent source coefficients. The final gas energy
after the \(\mathcal C\) step is instead obtained from the energy update in
Section~\ref{subsec:energy}.

\subsubsection{Picard iteration}

When the stopping time or charge-to-mass ratio depends on the evolving state, applying
the two-stage GIRK update to Equation~\eqref{eq:global_nonlinear} gives a nonlinear
system for the stage states. We solve this system by Picard iteration. At iteration
\(k\), \(\alpha_{a,i}^{(k)}\) and \(\xi_{a,i}^{(k)}\) are evaluated separately from
the two stage states and held fixed while the linear stage system is solved. The
resulting momenta and gas energies at the two stages are then used to reevaluate the
source coefficients. Both stage states are initialized to the state at the beginning
of the \(\mathcal C\) step, and \(\vct U^-\) remains fixed throughout the iteration.

Convergence is tested directly on the updated values of \(\alpha\) and \(\xi\). For
every dust species and both stages, we require
\begin{equation}
\begin{aligned}
  \left|\alpha_{a,i}^{(k+1)}-\alpha_{a,i}^{(k)}\right|
  &\leq \epsilon_\alpha\alpha_{a,i}^{(k)},\\
  \left|\xi_{a,i}^{(k+1)}-\xi_{a,i}^{(k)}\right|
  &\leq
  \epsilon_\xi\left|\xi_{a,i}^{(k)}\right|
  +\epsilon_{\xi,{\rm abs}},
  \qquad a=1,2,
\end{aligned}
\label{eq:picard_coefficient_convergence}
\end{equation}
where the second condition is applied when the magnetic field is nonzero and the
charge-to-mass ratio is state dependent. The absolute tolerance avoids an excessively
strict convergence requirement when \(\xi_{a,i}\) is close to zero. In contrast, for
the usual case of a positive finite stopping time, as in
Equations~\eqref{eq:epstein_stopping} and~\eqref{eq:epstein_baines_stopping}, the
drag rate \(\alpha_{a,i}=t_{{\rm s},a,i}^{-1}\) remains positive, so a relative
criterion is sufficient. If reevaluating the source coefficients changes the branch
selection, another Picard iteration is performed. By default, we use
\(\epsilon_\alpha=\epsilon_\xi=10^{-6}\) and
\(\epsilon_{\xi,{\rm abs}}=10^{-12}\). The iteration limit can be adjusted as needed
and is set to 20 by default. If the iteration limit is reached without convergence,
the final iterate is retained. If Picard iteration is disabled, the source
coefficients are held fixed at their values at the beginning of the \(\mathcal C\)
step. At each Picard iteration, the gas and dust stage rates are solved using the
reduction in Section~\ref{subsec:reduction}, with a cost of
\(\mathcal O(N_{\rm d})\). In the nonlinear tests of
Section~\ref{subsec:state_dependent_tests}, convergence requires only a few iterations
on average, so the total cost remains \(\mathcal O(N_{\rm d})\).

\subsection{Gas energy update}
\label{subsec:energy}

The \(\mathcal C\) step updates the gas energy through the mechanical work associated
with drag and the Lorentz force, together with the physical heating produced by
aerodynamic drag. Work done by external accelerations is included in the \(\mathcal H\)
update and is therefore not considered here.

From the momenta before and after one \(\mathcal C\) step, the gas and dust kinetic
energy changes are
\begin{align}
  \Delta E_{{\rm g},{\rm kin}}
  &=
  \frac{|\pg^+|^2-|\pg^-|^2}{2\rho_{\rm g}},\\
  \Delta E_{{\rm d},i,{\rm kin}}
  &=
  \frac{|\vct{p}_{{\rm d},i}^+|^2-|\vct{p}_{{\rm d},i}^-|^2}
  {2\rho_{{\rm d},i}}.
\end{align}

The Lorentz force has no separate physical heating term because
\((\vct{w}_i\times\hat{\vct b})\cdot\vct{w}_i=0\), where
\(\vct w_i=\vct v_{{\rm d},i}-\vg\). It can exchange kinetic energy between gas and
dust, but it does not dissipate their relative motion into heat. By contrast, for each
dust species, aerodynamic drag dissipates relative kinetic energy at the nonnegative
rate \(\alpha_i|\qrel_i|^2/\rho_{{\rm d},i}\).

The RK integration of the physical drag heating over the \(\mathcal C\) step gives
\begin{equation}
  \Delta E_{{\rm heat},{\rm RK}}^{\rm drag}
  =
  \Delta t_{\mathcal C}\sum_i\frac{1}{\rho_{{\rm d},i}}
  \left[
    b\alpha_{1,i}|\qrel_{1,i}|^2
    +(1-b)\alpha_{2,i}|\qrel_{2,i}|^2
  \right].
  \label{eq:physical_heat}
\end{equation}

Exact conservation of the total gas--dust energy over the discrete \(\mathcal C\) step
would require a compensating change in the gas internal energy,
\(\Delta E_{\rm cons}=-(\Delta E_{{\rm g},{\rm kin}}
+\sum_i\Delta E_{{\rm d},i,{\rm kin}})\).
We define the RK energy residual as the difference between this change and the RK
integration of the physical drag heating,
\begin{equation}
  \Delta E_{\rm res}^{\rm RK}
  =
  \Delta E_{\rm cons}
  -
  \Delta E_{{\rm heat},{\rm RK}}^{\rm drag}.
\end{equation}

We decompose the residual as
\(\Delta E_{\rm res}^{\rm RK} =\Delta E_{\rm res}^{\rm drag}+\Delta E_{\rm res}^{\Omega}\).
Here \(\Delta E_{\rm res}^{\rm drag}\) is obtained from the drag part of the stage
rates, while
\(\Delta E_{\rm res}^{\Omega} \equiv\Delta E_{\rm res}^{\rm RK}-\Delta E_{\rm res}^{\rm drag}\)
is the remaining gyrofrequency-dependent contribution and vanishes when all
\(\Omega_{{\rm L},i}=0\). Appendix~\ref{app:rk_energy} gives the explicit expressions
for this decomposition and shows that the RK energy residual vanishes identically for
GL4 and Midpoint, whereas it is generally nonzero for TP2025 and the stiff coefficients.
For TP2025, \(\Delta E_{\rm res}^{\rm RK}=\mathcal O(\Delta t_{\mathcal C}^3)\) as
\(\Delta t_{\mathcal C}\to0\), so including this residual does not reduce the
second-order accuracy of the source update for nonlinear source terms.

The gas total energy is updated as
\begin{equation}
\begin{aligned}
  E_{\rm g}^+
  ={}& E_{\rm g}^-
  + \Delta E_{{\rm g},{\rm kin}}\\
  &+\omega_{\rm drag}
  \left(\Delta E_{{\rm heat},{\rm RK}}^{\rm drag}
  +\Delta E_{\rm res}^{\rm drag}\right)
  +\omega_{\Omega}\Delta E_{\rm res}^{\Omega}.
\end{aligned}
\label{eq:gas_energy}
\end{equation}
Our default choice is \(\omega_{\rm drag}=1\) and \(\omega_{\Omega}=0\). Under this
choice, in the neutral limit, \(\Delta E_{\rm res}^{\Omega}=0\) and
\(\Delta E_{{\rm heat},{\rm RK}}^{\rm drag} +\Delta E_{\rm res}^{\rm drag}=\Delta E_{\rm cons}\),
so the conservative drag heating update is recovered. The gyrofrequency-dependent
residual \(\Delta E_{\rm res}^{\Omega}\) does not represent a separate physical heating
term, so our default choice \(\omega_{\Omega}=0\) does not deposit it as gas heat. In
the drag-free limit, it reduces to the correction that would compensate the numerical
change in gyromotion kinetic energy.

Other energy treatments may be appropriate depending on the application. Setting
\(\omega_{\rm drag}=\omega_{\Omega}=1\) deposits the full RK residual into the gas and
gives exact discrete conservation of the total gas--dust energy. Another option is to
omit the residual corrections and update the gas energy using only the RK integration of
the physical drag heating.

\subsection{Terminal drift and gyromotion diagnostics}
\label{subsec:relative_dynamics_diagnostics}

\begin{deluxetable*}{lll}[htb!]
\tablecaption{Diagnostics of the \(\mathcal C\mathcal H\mathcal C\)
Strang update for the RK coefficient sets used
in this work. GL4, Midpoint, and TP2025 are the resolved coefficient sets;
the final row gives the common stiff coefficients. The terminal drift
errors are given by \(\Wrel_{\rm fp}-{\Wrel}_*\); for the resolved
coefficient sets, the listed expressions give the leading behavior as
\(\Delta t\to0\), whereas \(\mathcal O(\Delta t^{-2})\) for the stiff
coefficients describes the large-timestep limit. For a drag-free
gyromotion mode with angular frequency \(\Omega\),
\(\theta=\Omega\Delta t\) is the exact phase advance over one timestep,
\(\delta a\) is the logarithmic amplitude error, and \(\delta\phi\) is the
phase error relative to \(\theta\).
\label{tab:combined_coeff_diagnostics}}
\tablehead{
\colhead{Coefficient set} &
\colhead{Terminal drift error \(\Wrel_{\rm fp}-{\Wrel}_*\)} &
\colhead{Drag-free gyromotion}
}
\startdata
GL4
&
\(\Delta t^2\mathcal{M}_{\rm rel}\vct{A}_{\rm rel}/24+
\mathcal{O}(\Delta t^3)\)
&
\(\delta a=0,\quad
\delta\phi=-\theta^5/11520+\mathcal{O}(\theta^7)\)
\\
Midpoint
&
\(\Delta t^2\mathcal{M}_{\rm rel}\vct{A}_{\rm rel}/16+
\mathcal{O}(\Delta t^3)\)
&
\(\delta a=0,\quad
\delta\phi=-\theta^3/48+\mathcal{O}(\theta^5)\)
\\
TP2025
&
\(\Delta t^2\mathcal{M}_{\rm rel}\vct{A}_{\rm rel}/24+
\mathcal{O}(\Delta t^3)\)
&
\(\delta a=-\theta^4/192+\mathcal{O}(\theta^6),\quad
\delta\phi=-\theta^5/720+\mathcal{O}(\theta^7)\)
\\
Stiff coefficients
&
\(\mathcal{O}(\Delta t^{-2})\)
&
\(\begin{gathered}
\delta a=\log[(1+\theta^2/4)/(1+\theta^4/4)],\quad
\delta\phi=
\operatorname{Arg}\!\left[
e^{-i\theta}
\left(\frac{1-i\theta/2}{1-i\theta-\theta^2/2}\right)^2
\right]
\end{gathered}\)
\\
\enddata
\end{deluxetable*}

We now quantify the behavior of the RK coefficient sets introduced in
Section~\ref{subsec:girk_source_update}. In the resolved regime, we use the amplitude
and phase errors of gyromotion as diagnostics. In the stiff limit, when external forces
act differently on the gas and dust, we examine whether the full
\(\mathcal C\mathcal H\mathcal C\) Strang update in Equation~\eqref{eq:strang}
approaches the terminal drift set by the drag--Lorentz balance.

For these diagnostics, we consider a uniform system with constant source coefficients
and external accelerations. We describe its motion using the center-of-mass velocity
\(\vct v_{\rm cm} =(\vg+\sum_i\epsilon_i\vct v_{{\rm d},i})/ (1+\sum_i\epsilon_i)\) and
the relative velocities \(\Wrel=(\vct w_1^T,\ldots,\vct w_{N_{\rm d}}^T)^T\), where
\(\vct w_i=\vct v_{{\rm d},i}-\vg\). During a \(\mathcal C\) update, momentum
conservation gives \(\dd\vct v_{\rm cm}/\dd t=0\), while the relative velocities obey
\(\dd\Wrel/\dd t=\mathcal M_{\rm rel}\Wrel\).

External accelerations are applied in \(\mathcal H\). Defining their difference by
\((\vct A_{\rm rel})_i =\vct a_{{\rm ext,d},i}-\vct a_{\rm ext,g}\), the relative
velocity equation for the full uniform system is
\begin{equation}
  \frac{\dd\Wrel}{\dd t}
  =
  \mathcal M_{\rm rel}\Wrel+\vct A_{\rm rel},
  \label{eq:Wrel_forced}
\end{equation}
where
\(\left(\mathcal M_{\rm rel}\Wrel\right)_i =\mtx T_i\vct w_i+\sum_k\epsilon_k\mtx T_k\vct w_k\).
If \(\mathcal M_{\rm rel}\) is invertible, setting \(\dd\Wrel/\dd t=0\) in
Equation~\eqref{eq:Wrel_forced} gives the steady solution
\begin{equation}
  {\Wrel}_*=-\mathcal{M}_{\rm rel}^{-1}\vct{A}_{\rm rel} ,
  \label{eq:terminal_general}
\end{equation}
which is the terminal drift. The component of the relative external acceleration
perpendicular to \(\vct B\) generally drives both a drift along that component and a
transverse drift perpendicular to both it and \(\vct B\). We refer to these as the
Pedersen-like and Hall-like components, respectively, by analogy with the corresponding
decomposition in weakly ionized plasma theory \citep[e.g.,][]{WardleNg1999}.

To determine the terminal drift selected by the numerical update, let \(R(z)\) denote
the RK stability function for one \(\mathcal C\) half-step, and define
\(\mtx Z=\Delta t_{\mathcal C}\mathcal M_{\rm rel}\). Applying one complete
\(\mathcal C\mathcal H\mathcal C\) Strang step gives
\begin{equation}
  \Wrel^{n+1}
  =
  R^2(\mtx{Z})\Wrel^n
  +
  \Delta t\,R(\mtx{Z})\vct{A}_{\rm rel}.
  \label{eq:split_forced_update}
\end{equation}
The fixed point of this update is
\begin{equation}
  \Wrel_{\rm fp}
  =
  \Delta t\left[\I-R^2(\mtx{Z})\right]^{-1}
  R(\mtx{Z})\vct{A}_{\rm rel}.
  \label{eq:combined_fixed_point}
\end{equation}
The difference \(\Wrel_{\rm fp}-{\Wrel}_*\) measures the error in the terminal drift
selected by the numerical update. Substituting the stability function for each RK
coefficient set into Equation~\eqref{eq:combined_fixed_point} gives the terminal drift
errors summarized in Table~\ref{tab:combined_coeff_diagnostics}.
Combining Equations~\eqref{eq:split_forced_update}
and~\eqref{eq:combined_fixed_point} gives
\(\Wrel^{n+1}-\Wrel_{\rm fp}=R^2(\mtx Z)(\Wrel^n-\Wrel_{\rm fp})\),
so unresolved drag relaxation transients are damped for the stiff coefficients
as \(R^2(\mtx Z)\to0\) in the large-timestep limit.

We next consider drag-free gyromotion with zero relative external acceleration,
\(\vct A_{\rm rel}=0\). In this limit, the gyromotion modes of \(\mathcal M_{\rm rel}\)
have purely imaginary eigenvalues. For one such mode, write
\(\mathcal M_{\rm rel}\vct e=i\Omega\vct e\) and define \(\theta=\Omega\Delta t\). One
\(\mathcal C\) half-step multiplies this mode by \(R(i\theta/2)\), so one complete
\(\mathcal C\mathcal H\mathcal C\) step multiplies it by \(R^2(i\theta/2)\). Since exact
gyromotion preserves the amplitude, \(\left|R^2(i\theta/2)\right|\) is the ratio of the
numerical amplitude to the exact amplitude after one timestep. We define the one-step
logarithmic amplitude error and phase error by
\begin{equation}
  \delta a = \log\left|R^2(i\theta/2)\right|,
  \qquad
  \delta\phi =
  \operatorname{Arg}\!\left[R^2(i\theta/2)e^{-i\theta}\right] .
\end{equation}
Here \(\operatorname{Arg}\) denotes the principal value in \((-\pi,\pi]\). The resulting
errors are listed in Table~\ref{tab:combined_coeff_diagnostics}. GL4 and Midpoint
preserve the gyromotion amplitude exactly, whereas TP2025 introduces numerical
damping. For the stiff coefficients, \(\left|R^2(i\theta/2)\right|\to0\) as
\(|\theta|\to\infty\), so unresolved gyromotion is suppressed in the large-timestep
limit.

\section{Results}\label{sec:tests}

We begin with spatially uniform problems that focus on the source update, testing its
drag damping, gyromotion, treatment of state-dependent source coefficients, and terminal
drift. We then consider a dusty Alfv\'en wave, a dusty Orszag--Tang vortex, and a
magnetized RDI in spatially varying dusty MHD flows. In these three tests, we use
second-order piecewise linear reconstruction (PLM) with the Sweby limiter for the gas
and dust states and the transverse magnetic field. The reconstructed gas states and
transverse magnetic field are supplied to the HLLD Riemann solver. Pressureless dust
fluxes are computed with the solver of \citet{HuangBai2022}. The magnetic field is
evolved with constrained transport using the B25a electromotive force (EMF) compute
scheme, the B25b EMF averaging scheme, and the extremum-preserving piecewise parabolic
method (PPM-EP) for reconstruction in the EMF calculation, following \citet{Kriel2026}.

Throughout this section, unless stated otherwise, quantities are given in the
dimensionless units defined in Section~\ref{subsec:dimensionless_units}. When
alternative resolved coefficient sets are compared, the scheme name refers to the
coefficient set used in the resolved branch; after the branch transition, all schemes
use the same stiff coefficients. When the stopping time or charge-to-mass ratio is state
dependent, we use the Picard iteration described in
Section~\ref{subsec:stopping_picard}.

\subsection{Two-species drag damping with widely separated stopping times}
\label{subsec:mixed_stopping_resolved}

We first test the \(\mathcal C\) update for pure aerodynamic drag. This test is based on
the dust damping problem of \citet{HuangBai2022}, but uses two dust species with widely
separated stopping times. We set \(\Omega_{{\rm L},i}=0\) for both species, so only
aerodynamic drag remains.

We use a uniform periodic box with side length \(L_{\rm box}=1\). The gas is initialized
with \(\rho_{\rm g}=1\), \(P_{\rm g}=1\), \(\gamma=1.4\), and \(\vg=(1,0,0)\). We use
two pressureless dust species with \(\rho_{{\rm d},1}=\rho_{{\rm d},2}=1\),
\(\vct{v}_{{\rm d},1}=(2,0,0)\), and \(\vct{v}_{{\rm d},2}=(0.5,0,0)\). Their stopping
times are \(t_{{\rm s},1}=0.2\) and \(t_{{\rm s},2}=2\times10^{-3}\). We evolve the
system to \(t=2.0\) with a fixed timestep \(\Delta t=0.1\). We compute the analytic
reference solution following the same procedure as in \citet{HuangBai2022}. Both the
numerical and analytic solutions use \(\omega_{\rm drag}=1\).

With \(\Omega_{{\rm L},i}=0\), Equation~\eqref{eq:tauDL} reduces to
\(\tau_{\rm DL}=\max_i t_{{\rm s},i}=0.2\), so the resolved branch is selected because
\(\Delta t<\tau_{\rm DL}\). The stopping time of the second dust species is nevertheless
unresolved, with \(\Delta t/t_{{\rm s},2}=50\). Thus the test compares the resolved
coefficient sets when the branch is selected by \(t_{{\rm s},1}\), even though the
shorter stopping time \(t_{{\rm s},2}\) is unresolved.

\begin{figure}[htb!]
\centering
\includegraphics[width=\columnwidth]{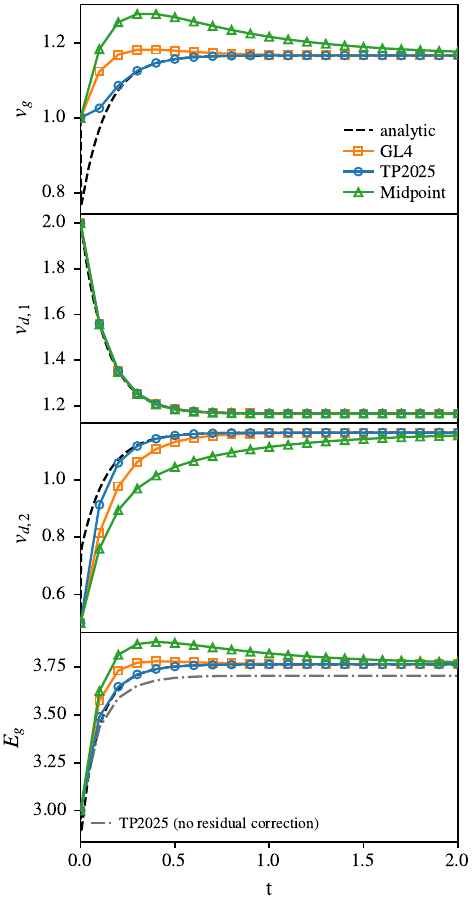}
\caption{Time evolution for pure drag with two dust species having
widely separated stopping times. Black dashed curves show the analytic
solution, and colored curves with symbols show the numerical
solutions using the three resolved coefficient sets. In
the bottom panel, the gray dash-dotted curve uses the TP2025 momentum
update but omits \(\Delta E_{\rm res}^{\rm drag}\) from the gas energy
update.}
\label{fig:mixed_stopping_resolved}
\end{figure}

Figure~\ref{fig:mixed_stopping_resolved} compares the numerical and analytic evolution
of the gas velocity, the two dust velocities, and the gas energy. The \(v_{{\rm d},1}\)
evolution is reproduced well by all three resolved coefficient sets, while the gas
velocity, \(v_{{\rm d},2}\), and gas energy show clearer differences among the schemes.
TP2025 follows the analytic solution most closely, GL4 is intermediate, and Midpoint
shows the largest transient deviations. These differences arise mainly from the
unresolved transient associated with the shorter stopping time, which TP2025 suppresses
most strongly. All three solutions nevertheless eventually approach the same equilibrium
velocity. The gas energy panel also shows a TP2025 calculation with the same momentum
update but with \(\Delta E_{\rm res}^{\rm drag}\) omitted from the gas energy update.
The resulting energy deficit is the difference between the RK integration of the physical
drag heating and the conservative energy change implied by the discrete momentum update.
Including \(\Delta E_{\rm res}^{\rm drag}\) restores conservation of the total gas--dust
energy, as described in Section~\ref{subsec:energy}. No separate uncorrected curves
are shown for GL4 or Midpoint because their RK energy residuals vanish identically,
making the corrected and uncorrected curves coincide.

We next vary the full timestep from \(10^{-4}\) to \(10^3\). To compare the numerical
and analytic velocity evolution, we define
\begin{equation}
  e_v
  =
  \frac{1}{N_{\rm step}}
  \sum_{k=1}^{N_{\rm step}}
  \sum_v
  \left|
    \frac{v^{\rm num}(t_k)-v^{\rm exact}(t_k)}
         {v^{\rm exact}(t_k)}
  \right|.
  \label{eq:mixed_stopping_velocity_error}
\end{equation}
The inner sum runs over the gas and dust velocities, \(v_{\rm g}\), \(v_{{\rm d},1}\),
and \(v_{{\rm d},2}\). For \(\Delta t<t_{{\rm s},1}\), we use
\(N_{\rm step}=\lceil t_{{\rm s},1}/\Delta t\rceil\), so that each calculation samples
the transient evolution for approximately \(t_{{\rm s},1}\). For
\(\Delta t\geq t_{{\rm s},1}\), we instead take one step and evaluate the error at
\(t=\Delta t\) to examine the large-timestep behavior of the stiff update.

\begin{figure}[htb!]
\centering
\includegraphics[width=\columnwidth]{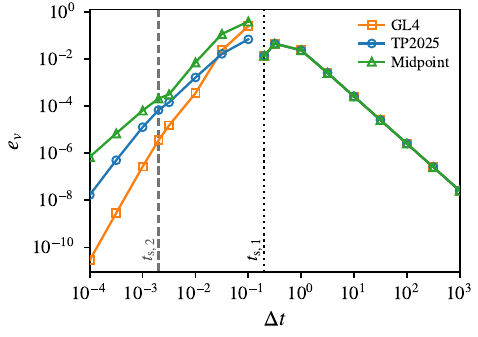}
\caption{Velocity error \(e_v\) from
Equation~\eqref{eq:mixed_stopping_velocity_error} as a function of the
full timestep. The gray dashed vertical line marks
\(\Delta t=t_{{\rm s},2}\), where the shorter stopping time becomes
unresolved, and the black dotted line marks the branch transition at
\(\Delta t=t_{{\rm s},1}=\tau_{\rm DL}\).}
\label{fig:mixed_stopping_timestep_sweep}
\end{figure}

When both stopping times are resolved, the errors in
Figure~\ref{fig:mixed_stopping_timestep_sweep} recover the expected fourth-, third-, and
second-order convergence of GL4, TP2025, and Midpoint, respectively. For
\(t_{{\rm s},2}<\Delta t<t_{{\rm s},1}\), the shorter stopping time is unresolved, but
the resolved branch is still selected. As the timestep increases within this range,
TP2025 suppresses the unresolved transient associated with \(t_{{\rm s},2}\) more
strongly, and its error becomes smaller relative to GL4 and Midpoint. At
\(\Delta t=0.1\), the timestep used in Figure~\ref{fig:mixed_stopping_resolved}, TP2025
gives the smallest error, consistent with the time evolution shown there. At
\(\Delta t=t_{{\rm s},1}=\tau_{\rm DL}\), all three calculations switch to the common
stiff coefficient set and their curves coincide. At still larger timesteps, the common
stiff update increasingly suppresses the unresolved transient and the velocity error
decreases toward the equilibrium limit.

These results show the expected high-order accuracy of GL4 when both stopping times are
resolved. When the shorter stopping time becomes unresolved, TP2025 gives smaller errors
as the timestep increases toward the branch transition. We therefore use TP2025 as the
default resolved coefficient set without MHD. Section~\ref{subsec:gyro_tests} then shows
why accurate gyromotion favors GL4 as the default resolved coefficient set when MHD is
included.

\subsection{Charged dust damping and gyromotion}
\label{subsec:gyro_tests}

We next consider the damping and gyromotion tests in \citet[\S3.1]{Moseley2023}. The
calculations use one gas fluid and one dust fluid in a periodic box. The gas is
isothermal with \(\rho_{\rm g}=1\) and \(c_{\rm s}=1\), the dust has \(\rho_{\rm d}=1\),
and the center-of-mass velocity is zero. The initial dust--gas relative velocity is
\(\vct w=\vd-\vg=(10,0,0)\), corresponding to \(v_{{\rm g},x}=-5\) and
\(v_{{\rm d},x}=5\).

The test includes three cases: pure drag damping, drag-free gyromotion, and damped
gyromotion. When drag is included, we use Equation~\eqref{eq:epstein_baines_stopping}
for the stopping time, with \(t_{\rm s,0}=1\). In the magnetized cases,
\(\vct B=5\hat{\vct z}\) and the dimensionless charge-to-mass ratio is \(\xi=1\), giving
\(\Omega_{\rm L}=5\). For the drag-free case, the corresponding gyromotion frequency
defined in Section~\ref{subsec:relative_dynamics_diagnostics} is
\(\Omega=(1+\epsilon)\Omega_{\rm L}=10\). All runs use \(\Delta t=0.1\) and are evolved
to \(t=2\). We plot the normalized relative velocity \(w_x/w_0\), where \(w_0=10\) is
the initial relative speed, and compare with the corresponding analytic solutions
given in \citet{Moseley2023}.

\begin{figure*}[t!]
\centering
\includegraphics[width=\textwidth]{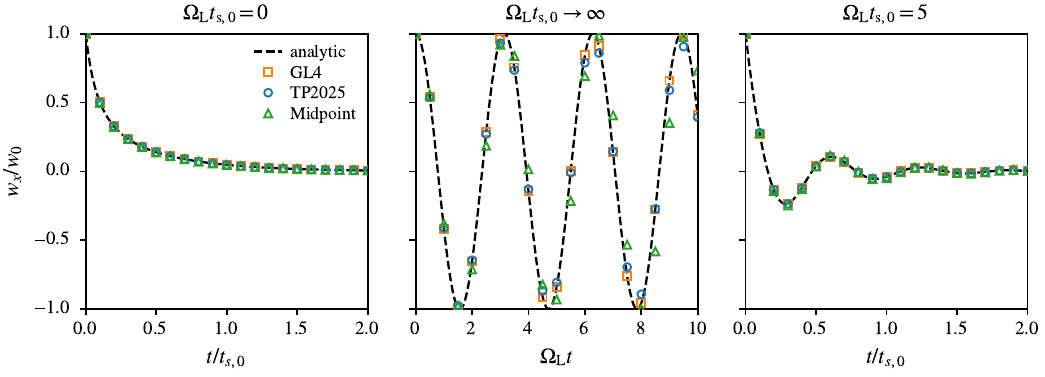}
\caption{Normalized dust--gas relative velocity in pure drag damping (left),
drag-free gyromotion (middle), and damped gyromotion (right) following
\citet[\S3.1]{Moseley2023}. Dashed black curves show the analytic
solutions, and colored symbols show the numerical solutions using the three
resolved coefficient sets.}
\label{fig:damped_gyromotion_tests}
\end{figure*}

Figure~\ref{fig:damped_gyromotion_tests} shows that all three resolved coefficient sets
reproduce the drag damping and damped gyromotion well. The drag-free case more clearly
separates their amplitude and phase errors. TP2025 gradually loses amplitude, while
Midpoint preserves the amplitude but accumulates a phase error, consistent with the
corresponding result in \citet{Moseley2023}, whose Lorentz update uses an implicit
midpoint scheme. GL4 gives the closest agreement with the analytic solution.

\begin{figure}[htb!]
\centering
\includegraphics[width=\columnwidth]{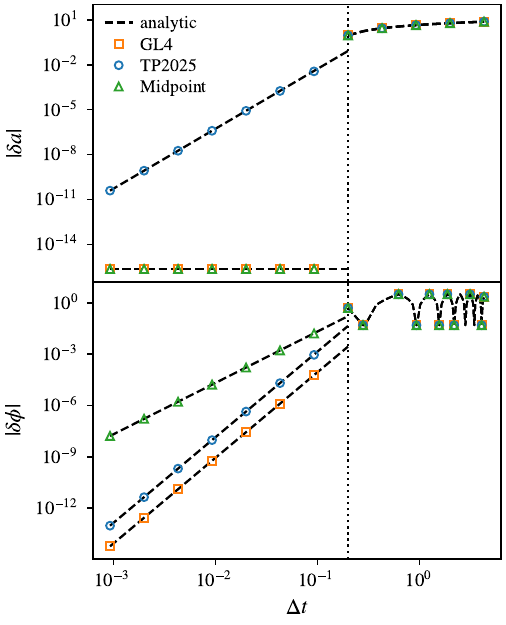}
\caption{Absolute one-step logarithmic amplitude error \(|\delta a|\)
(upper) and phase error \(|\delta\phi|\) (lower) of drag-free gyromotion
as functions of the full timestep. Colored symbols show the measured errors,
and black dashed curves show the analytic predictions from
Section~\ref{subsec:relative_dynamics_diagnostics} and
Table~\ref{tab:combined_coeff_diagnostics}. The vertical dotted line
marks the branch transition at
\(\Delta t=|\Omega_{\rm L}|^{-1}\).}
\label{fig:gyromotion_diagnostics}
\end{figure}

We next examine the drag-free problem in the middle panel of
Figure~\ref{fig:damped_gyromotion_tests} more quantitatively.
Figure~\ref{fig:gyromotion_diagnostics} shows the one-step logarithmic amplitude and
phase errors as functions of the full timestep \(\Delta t\), spanning both the resolved
and stiff branches. The measured errors are compared with the predictions in
Section~\ref{subsec:relative_dynamics_diagnostics} and
Table~\ref{tab:combined_coeff_diagnostics}.

In the resolved branch, the TP2025 logarithmic amplitude error follows the predicted
\(\delta a=-\theta^4/192+\mathcal O(\theta^6)\), and hence scales as \(\Delta t^4\),
while GL4 and Midpoint preserve the amplitude to roundoff. Midpoint has the largest
phase error because its leading phase error is lower order, whereas GL4 has the smallest
phase error. These results support using GL4 as the default resolved coefficient set
when MHD is included. In the stiff branch, all three schemes use the same stiff
coefficients, and the measured amplitude and phase errors follow the analytic
predictions in Table~\ref{tab:combined_coeff_diagnostics}.

\subsection{State-dependent stopping time and charge-to-mass ratio}
\label{subsec:state_dependent_tests}

We now extend the damped gyromotion problem in the right panel of
Figure~\ref{fig:damped_gyromotion_tests} to test the nonlinear source update when the
source coefficients vary with the evolving state. In this problem, the gas density and
sound speed remain fixed, so Equation~\eqref{eq:epstein_baines_stopping} makes the
stopping time depend only on the magnitude of the dust--gas relative velocity. Defining
\(W\equiv|\vct w|\), we denote this dependence by \(t_{\rm s}(W)\).

We consider two cases. In both, the stopping time varies as \(t_{\rm s}(W)\). In the
first, the dimensionless charge-to-mass ratio is fixed at \(\xi=1\). In the second, we
prescribe a state-dependent charge-to-mass ratio to test the nonlinear source update
as the charge changes sign:
\begin{equation}
  \xi(W)=\frac{W}{w_0}-\xi_{\rm off},
  \label{eq:dynamic_test_charge}
\end{equation}
with \(\xi_{\rm off}=0.95\). The dimensionless charge-to-mass ratio therefore starts at
\(\xi=0.05\) and becomes negative as the relative velocity is damped.

This prescription also allows an analytic solution. Since the Lorentz force remains
perpendicular to \(\vct w\), varying \(\xi\) changes the gyromotion phase but not the
magnitude \(W\). Both cases therefore have the same \(W(t)\), given by the analytic
solution of \citet[\S3.1]{Moseley2023} for damped gyromotion with the stopping time in
Equation~\eqref{eq:epstein_baines_stopping}. Writing \(w_x-iw_y=W e^{i\phi}\), the phase
satisfies \(\dd\phi/\dd t=(1+\epsilon)|\vct B|\,\xi[W(t)]\). With
\(\alpha_0=t_{{\rm s},0}^{-1}\) and \(\eta=9\pi\gamma/128\), integration gives
\begin{align}
  \phi(t)
  ={}&
  \frac{|\vct B|c_s}{\alpha_0w_0\sqrt{\eta}}
  \left[
    \sinh^{-1}\!\left(\frac{\sqrt{\eta}w_0}{c_s}\right)
    -\sinh^{-1}\!\left(\frac{\sqrt{\eta}W(t)}{c_s}\right)
  \right]
  \notag\\
  &-(1+\epsilon)|\vct B|\xi_{\rm off}t.
  \label{eq:dynamic_test_phase}
\end{align}

For both cases, we use the GL4 coefficients and compare the Picard iteration with
calculations using frozen source coefficients. In the latter, the source coefficients
are evaluated from the state at the beginning of each source step and held fixed at both
stages.

\begin{figure}[htb!]
\centering
\includegraphics[width=\columnwidth]{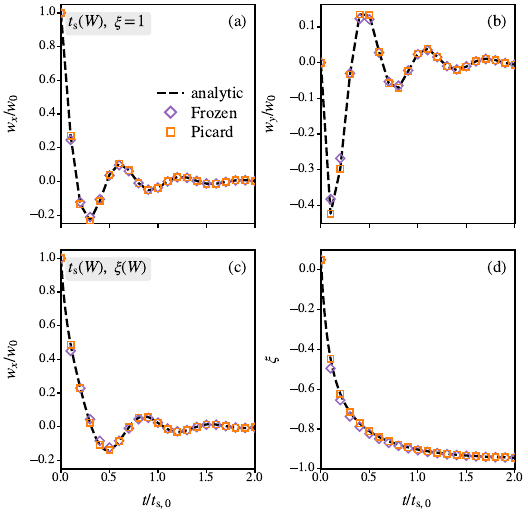}
\caption{Time evolution for two cases of damped gyromotion with
state-dependent source coefficients. The first uses a state-dependent
stopping time and a constant dimensionless charge-to-mass ratio
(\(t_{\rm s}(W)\), \(\xi=1\); top row), while in the second both are
state dependent (\(t_{\rm s}(W)\), \(\xi(W)\); bottom row). Black
dashed curves show the analytic solutions, purple diamonds show the
results with frozen source coefficients, and orange squares show the results
with Picard iteration.}
\label{fig:dynamic_coefficient_trajectories}
\end{figure}

Figure~\ref{fig:dynamic_coefficient_trajectories} compares the numerical and analytic
time evolution. For the case with fixed \(\xi=1\), we plot the normalized dust--gas
relative velocity components \(w_x/w_0\) and \(w_y/w_0\). For the case with
state-dependent \(\xi\), we plot \(w_x/w_0\) together with the dimensionless
charge-to-mass ratio \(\xi\). We use \(\Delta t/t_{{\rm s},0}=0.1\), evolve the system
to \(t/t_{{\rm s},0}=2\), and use the default Picard tolerances
\(\epsilon_\alpha=\epsilon_\xi=10^{-6}\) and \(\epsilon_{\xi,{\rm abs}}=10^{-12}\).

In both cases, updating the source coefficients with Picard iteration gives close
agreement with the analytic relative velocity, whereas freezing the source coefficients
produces visible errors. In the case with state-dependent \(\xi\), the result with
Picard iteration also reproduces the analytic evolution of \(\xi\) as it crosses zero
and becomes negative. For the calculations shown in
Figure~\ref{fig:dynamic_coefficient_trajectories}, the mean numbers of Picard iterations
per source half-step are \(3.18\) for the case with fixed \(\xi=1\) and \(3.23\) for the
case with state-dependent \(\xi=\xi(W)\). All source updates in these tests converged
without reaching the default limit of 20 iterations. The modest iteration counts are
consistent with the relatively mild state dependence in these tests. For example, the
stopping time \(t_{\rm s}(W)\) varies smoothly with \(W\).

\begin{figure}[htb!]
\centering
\includegraphics[width=\columnwidth]{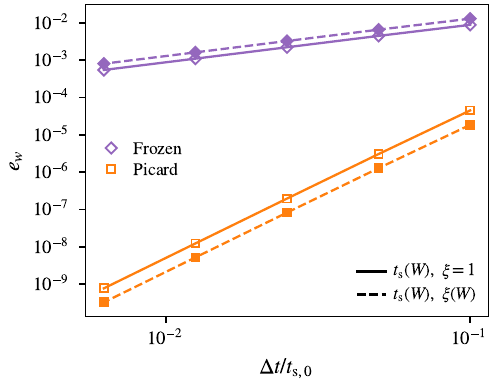}
\caption{Error \(e_w\), defined by
Equation~\eqref{eq:dynamic_gyro_error}, as a function of the normalized
timestep \(\Delta t/t_{{\rm s},0}\). Purple diamonds show the results
with frozen source coefficients and orange squares show the results with Picard
iteration. Solid lines with open markers show the case with fixed
\(\xi=1\), while dashed lines with filled markers show the case with
state-dependent \(\xi=\xi(W)\). All calculations use the GL4
coefficients.}
\label{fig:dynamic_coefficient_errors}
\end{figure}

We next vary the full timestep \(\Delta t\) and measure the error in the dust--gas
relative velocity as
\begin{equation}
  e_w
  =
  \frac{1}{N_{\rm step}}
  \sum_{k=1}^{N_{\rm step}}
  \frac{
    \left|\vct w_{\rm num}(t_k)-\vct w_{\rm exact}(t_k)\right|
  }{w_0}.
  \label{eq:dynamic_gyro_error}
\end{equation}
Here \(N_{\rm step}\) is the number of timesteps used to evolve the system to
\(t=2t_{{\rm s},0}\). Figure~\ref{fig:dynamic_coefficient_errors} shows \(e_w\) as a
function of \(\Delta t/t_{{\rm s},0}\). The Picard calculations in this figure use
\(\epsilon_\alpha=\epsilon_\xi=\epsilon_{\xi,{\rm abs}}=10^{-12}\), so that the
iteration tolerance does not limit the measured temporal error.

Figure~\ref{fig:dynamic_coefficient_errors} shows that calculations with frozen source
coefficients are approximately first order in both cases, while the Picard iteration
recovers the fourth-order convergence of GL4. Thus, evaluating the state-dependent
source coefficients at their corresponding RK stage states is necessary to
preserve this convergence order.

\subsection{Terminal drift under differential external forces}
\label{subsec:forced_relative_drift}

We next consider terminal drift produced by different external accelerations on the gas
and dust, with the relative acceleration perpendicular to \(\vct B\). This setup tests
whether the drag--Lorentz update recovers the resulting Pedersen-like component along
\(\vct a_{\rm rel}\) and Hall-like component perpendicular to both \(\vct a_{\rm rel}\)
and \(\vct B\).

We use one gas fluid and one dust fluid in a periodic box of side length
\(L_{\rm box}=1\). The gas is isothermal with \(\rho_{\rm g}=1\) and \(c_{\rm s}=1\),
while the dust has \(\rho_{\rm d}=1\), giving \(\epsilon=1\). We set \(t_{\rm s}=1\),
and hence \(\alpha=1\), together with \(\vct{B}=\hat{\vct z}\) and the dimensionless
charge-to-mass ratio \(\xi=1\), giving \(\Omega_{\rm L}=1\). Both gas and dust are
initialized at rest. We apply \(\vct a_{\rm ext,g}=\hat{\vct x}\) to the gas and set
\(\vct a_{\rm ext,d}=0\), giving \(\vct a_{\rm rel}=-\hat{\vct x}\), perpendicular to
\(\vct B\).

For a single dust species, Equation~\eqref{eq:Wrel_forced} reduces to the following
equation for the dust--gas relative velocity \(\vct w=\vd-\vg\),
\begin{equation}
  \frac{\dd\vct{w}}{\dd t}
  =
  -(1+\epsilon)\alpha\vct{w}
  +(1+\epsilon)\Omega_{\rm L}
  \vct{w}\times\hat{\vct b}
  +\vct{a}_{\rm rel}.
\end{equation}
For the parameters above, setting \(\dd\vct w/\dd t=0\), equivalently using
Equation~\eqref{eq:terminal_general}, gives the exact terminal drift \(w_{*,x}=-1/4\)
and \(w_{*,y}=1/4\). The \(x\)-component is the Pedersen-like component along
\(\vct a_{\rm rel}\), while the \(y\)-component is the Hall-like component perpendicular
to both \(\vct a_{\rm rel}\) and \(\vct B\).

\begin{figure}[htb!]
\centering
\includegraphics[width=\columnwidth]{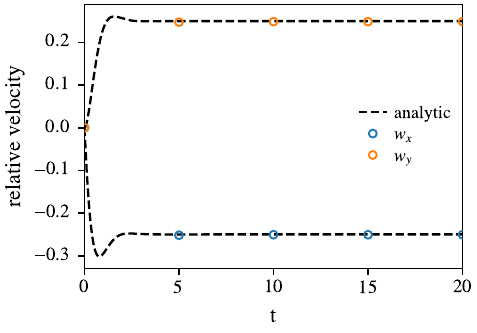}
\caption{Time evolution of the Pedersen-like (\(w_x\)) and Hall-like
(\(w_y\)) components of the dust--gas relative velocity in the
terminal drift test. Black dashed curves show the analytic solution, and
symbols show the numerical solution.}
\label{fig:hall_pedersen_drift}
\end{figure}

Figure~\ref{fig:hall_pedersen_drift} compares the analytic solution with a numerical
calculation evolved to \(t=20\) using \(\Delta t=5\), which selects the stiff branch.
The numerical solution approaches the analytic terminal values of both the Pedersen-like
and Hall-like components.

\begin{figure}[htb!]
\centering
\includegraphics[width=\columnwidth]{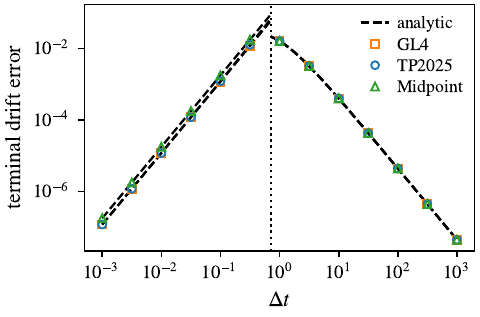}
\caption{Terminal drift error as a function of timestep for the problem in
Figure~\ref{fig:hall_pedersen_drift}. Colored symbols show the measured
error \(|\vct{w}_{\rm final}-\vct{w}_*|\), and black dashed curves show the
analytic fixed-point offset \(|\vct{w}_{\rm fp}-\vct{w}_*|\) from
Equation~\eqref{eq:combined_fixed_point}. The vertical dotted line marks the
branch transition at \(\Delta t=\tau_{\rm DL}=1/\sqrt{2}\).}
\label{fig:forced_diagnostics}
\end{figure}

Figure~\ref{fig:forced_diagnostics} compares the final numerical error
\(|\vct{w}_{\rm final}-\vct{w}_*|\) with the offset of the discrete fixed point,
\(|\vct{w}_{\rm fp}-\vct{w}_*|\), over timesteps spanning both the resolved and stiff
branches. Here \(\vct{w}_{\rm fp}\) is the fixed point defined by
Equation~\eqref{eq:combined_fixed_point}. Each calculation uses
\(N_{\rm step}=\max\{\lceil20/\Delta t\rceil,2\}\) full timesteps, so that the evolution
extends to at least \(t=20\) and \(\vct{w}_{\rm final}\) has time to approach
\(\vct{w}_{\rm fp}\).

As predicted in Section~\ref{subsec:relative_dynamics_diagnostics} and
Table~\ref{tab:combined_coeff_diagnostics}, the measured terminal drift error
\(|\vct{w}_{\rm final}-\vct{w}_*|\) follows the \(\mathcal{O}(\Delta t^2)\) fixed-point
offset \(|\vct{w}_{\rm fp}-\vct{w}_*|\) in the resolved branch. On the stiff branch,
both quantities approach the predicted \(\mathcal{O}(\Delta t^{-2})\) scaling at large
timesteps. The agreement across both branches shows that the discrete solution recovers
the Pedersen-like and Hall-like components of the terminal drift set by the
drag--Lorentz balance. This is also why we advance drag and Lorentz forces together
rather than in separate source substeps. In the stiff limit, a split treatment such as
that used by \citet{Moseley2023} can instead converge to a different terminal drift.

\subsection{Dusty Alfv\'en wave}
\label{subsec:dusty_alfven_wave}

\begin{figure*}[t!]
\centering
\includegraphics[width=\textwidth]{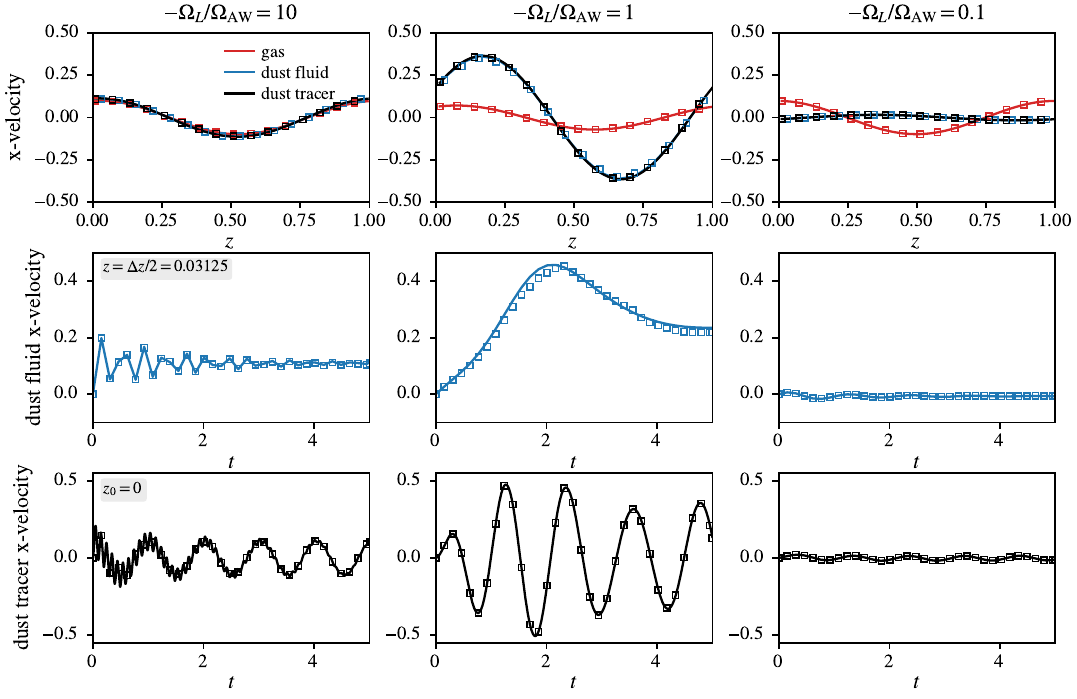}
\caption{Gas, dust fluid, and tracer \(x\)-velocity profiles at \(t=5\) (top row),
time evolution of the dust fluid \(x\)-velocity at the first cell center
\(z=\Delta z/2=0.03125\) (middle row), and time evolution of the \(x\)-velocity of a
tracer initially at \(z_0=0\) (bottom row) in the dusty Alfv\'en wave test. From left
to right, the columns show initial frequency ratios
\(-\Omega_{\rm L}/\Omega_{\rm AW}=10\), \(1\), and \(0.1\) at fixed dust-to-gas mass
ratio \(\epsilon=0.01\) and stopping time \(\ts=10/\Omega_{\rm AW}\). Curves show
the high-accuracy reference solution. Squares show dust fluid velocities taken directly
from the simulation, gas velocities reconstructed at tracer positions, and tracer
velocities calculated in post-processing.}
\label{fig:dusty_alfven_omega}
\end{figure*}

We next consider the dusty Alfv\'en wave test of \citet[\S3.3]{Moseley2023}. Unlike the
preceding spatially uniform tests, this problem includes spatially varying MHD evolution
and tests charged dust motion in an Alfv\'en wave.
We use a periodic cubic box with \(L_{\rm box}=1\), resolved by a uniform
\(8\times8\times16\) grid. The gas is isothermal with initial \(P=\rho_{\rm g}=1\).
We initialize a circularly polarized Alfv\'en wave with \(k=2\pi/L_{\rm box}=2\pi\),
\(\vg=\vct B=(0.1\cos kz,\,0.1\sin kz,\,1)\), and \(\vd=(0,0,1)\). The Alfv\'en
speed is \(v_{\rm A}=1\), giving a wave angular frequency
\(\Omega_{\rm AW}=k v_{\rm A}=2\pi\). We fix the dust-to-gas mass ratio at
\(\epsilon=0.01\) and the stopping time at \(\ts=10/\Omega_{\rm AW}\), and use
constant charge-to-mass ratios giving initial frequency ratios
\(-\Omega_{\rm L}/\Omega_{\rm AW}=10\), \(1\), and \(0.1\). We evolve each case to
\(t=5\) with a CFL number of \(0.3\).

In our calculation, dust is represented by an Eulerian fluid, while \citet{Moseley2023}
use MHD-PIC superparticles. To follow individual dust trajectories as in
their test, we construct a passive dust tracer diagnostic in post-processing.
Figure~\ref{fig:dusty_alfven_omega} shows gas, dust fluid, and tracer \(x\)-velocity
profiles at \(t=5\) in the top row. The middle row records the dust fluid velocity
at a fixed position, whereas the bottom row follows the velocity along a tracer
trajectory.

The tracers are initially distributed uniformly along \(z\), with the same initial
velocity as the dust fluid. At each stored time, we extract the fundamental circularly
polarized modes of the transverse gas velocity and magnetic field, together with their
mean longitudinal components. We interpolate these quantities linearly in time and
reconstruct the fields at each tracer position. We then integrate the tracer positions
and velocities under aerodynamic drag and the Lorentz force. The tracers do not affect
the gas or magnetic field; dust back-reaction is already included in the multifluid
simulation.

For comparison, we also construct a high-accuracy reference solution. Because the
transverse components of the gas and dust velocities and magnetic field retain a
single circularly polarized mode, their evolution can be reduced to ordinary
differential equations for the mode amplitudes and longitudinal gas and dust
velocities. We integrate these equations to obtain the reference gas and dust
velocity fields. We then integrate the same initial tracer ensemble through the
reference gas velocity and magnetic field to obtain the tracer reference.

\begin{figure*}[t!]
\centering
\includegraphics[width=\textwidth]{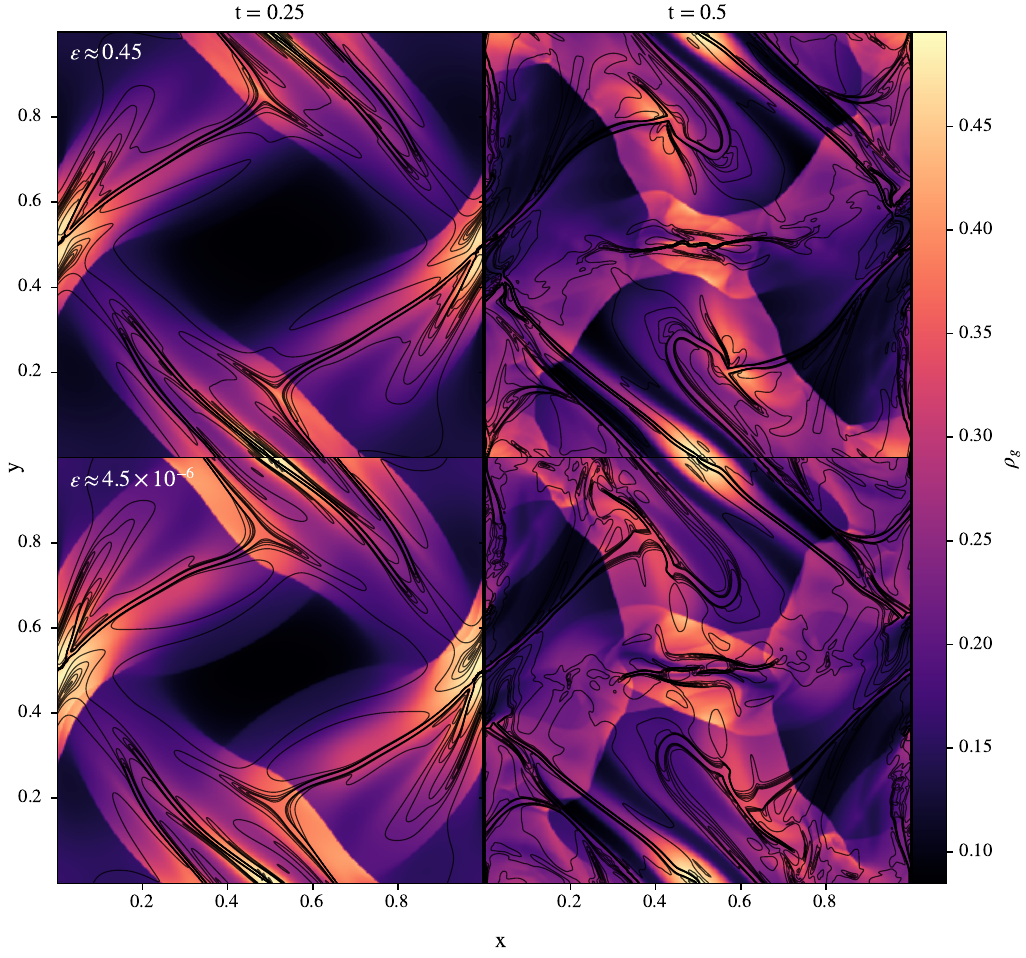}
\caption{Gas density maps with dust overdensity contours in the dusty Orszag--Tang
vortex following \citet[\S3.4]{Moseley2023}. The black contours show
\(\rho_{\rm d}/\langle\rho_{\rm d}\rangle\) from \(0.1\) to \(3.1\) in intervals of
\(0.5\), for initial dust-to-gas mass ratios \(\epsilon\approx0.45\) (top row) and
\(\epsilon\approx4.5\times10^{-6}\) (bottom row).}
\label{fig:dusty_orszag_tang_panels}
\end{figure*}

The top row of Figure~\ref{fig:dusty_alfven_omega} shows that the dust fluid and tracer
profiles nearly coincide at \(t=5\). The resonant case in the middle column,
\(-\Omega_{\rm L}/\Omega_{\rm AW}=1\), produces the largest transverse velocities in
both dust diagnostics. Across all three columns, the numerical profiles and time
histories agree closely with their respective references in amplitude and phase.
We also repeated the resonant test of \citet[\S3.3]{Moseley2023} for different values
of \(\epsilon\) and found similarly close agreement between the tracer diagnostic
and the high-accuracy reference (not shown).

\subsection{Dusty Orszag--Tang vortex}
\label{subsec:dusty_orszag_tang}

We next consider the dusty Orszag--Tang vortex test of \citet[\S3.4]{Moseley2023}. This
problem extends the standard compressible Orszag--Tang vortex by adding a uniformly
distributed charged dust component that is initially comoving with the gas. It therefore
tests the drag--Lorentz update in a multidimensional MHD flow containing shocks, where
dust is concentrated by the shocks and can back-react on the gas.

We use a periodic box with \(L_{\rm box}=1\) and the same compressible Orszag--Tang gas
initial conditions, with \(\gamma=5/3\), \(\rho_{\rm g,0}=25/(36\pi)\), and
\(P_0=5/(12\pi)\). The dust has a constant stopping time \(t_{\rm s}=0.1\) and a
dimensionless charge-to-mass ratio \(\xi=100\). We use initial dust densities
\(\rho_{\rm d,0}=10^{-1}\) and \(10^{-6}\), corresponding to initial dust-to-gas mass
ratios \(\epsilon\approx0.45\) and \(4.5\times10^{-6}\), respectively. Both calculations
use an in-plane resolution of \(512^2\) cells and eight cells along the invariant \(z\)
direction. We use a CFL number of \(0.3\).

Figure~\ref{fig:dusty_orszag_tang_panels} shows the gas density and dust overdensity at
\(t=0.25\) and \(t=0.5\). The color scale gives the gas density, while the black
contours trace the dust overdensity \(\rho_{\rm d}/\langle\rho_{\rm d}\rangle\). After
the shocks form, the dust concentrates into narrow sheets that are stretched by the
vortical flow. For \(\epsilon\approx4.5\times10^{-6}\), dust back-reaction is negligible
and the gas remains close to the dust-free Orszag--Tang evolution. For
\(\epsilon\approx0.45\), the dust modifies the gas dynamics, and the shock evolution is
slightly delayed relative to the low-\(\epsilon\) case.

\begin{figure}[htb!]
\centering
\includegraphics[width=\columnwidth]{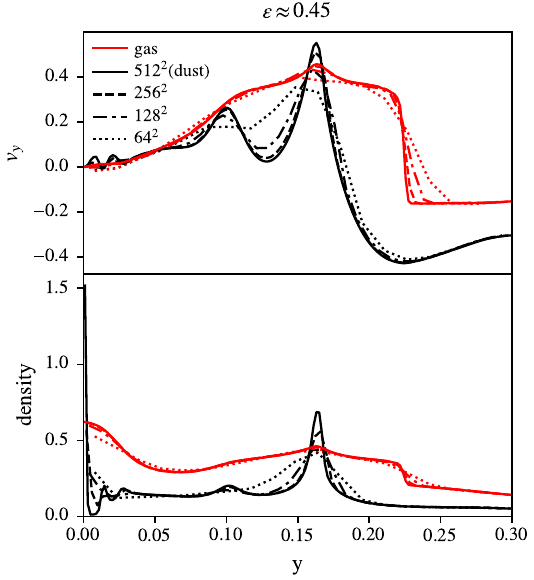}
\caption{\(x=0.5\) profiles of the dusty Orszag--Tang test at \(t=0.25\) for
\(\epsilon\approx0.45\): gas and dust \(y\)-velocity profiles (upper), and gas and dust
density profiles (lower). The curves show in-plane resolutions from \(64^2\) to
\(512^2\).}
\label{fig:dusty_orszag_tang_profiles}
\end{figure}

To examine the resolution dependence, we repeat the calculation with
\(\epsilon\approx0.45\) at in-plane resolutions from \(64^2\) to \(512^2\).
Figure~\ref{fig:dusty_orszag_tang_profiles} shows the \(x=0.5\) profiles at \(t=0.25\).
The upper panel compares the \(y\)-velocity profiles of the gas and dust, while the
lower panel compares their density profiles.

Across the four resolutions, the positions of the main velocity extrema and density
peaks are similar. The gas curves change relatively little as the grid is refined,
whereas the dust velocity and density show clearer resolution dependence. In the dust
velocity, the two peaks near \(y\approx0.10\) and \(0.16\) and the trough between them
continue to change noticeably with resolution. In this test, \citet{Moseley2023}
identify the post-shock oscillations in the dust velocity and density as
``dust-cyclotron'' waves. The dust density peak near \(y\approx0.16\) becomes narrower
and higher as the resolution increases, indicating that its shape and amplitude are not
yet fully converged. The narrow enhancement near the periodic boundary is also
especially prominent at \(512^2\).

\subsection{Magnetized resonant drag instability}
\label{subsec:magnetized_rdi}

We finally consider a magnetized RDI. RDIs of charged dust
in gas obeying ideal MHD have been studied analytically \citep{HopkinsSquire2018} and
numerically \citep{HopkinsSquireSeligman2020}. Here we adopt the test of
\citet[\S3.5]{Moseley2023}, which in turn reproduces the magnetized RDI simulation of
\citet{Seligman2019}. A sustained drift between the dust and gas in this problem excites
MHD waves through dust back-reaction.

Following \citet{Moseley2023}, we use a periodic box with \(L_{\rm box}=1\), resolved by
\(128^3\) cells, and one dust fluid. We use a CFL number of \(0.25\). The gas is
isothermal with \(\rho_{\rm g}=c_s=1\), and the dust-to-gas mass ratio is
\(\epsilon=0.01\). We use the approximation to the Epstein--Baines drag law in
Equation~\eqref{eq:epstein_baines_stopping}, setting \(a\rho_{\rm gr}=5\) in that
expression, and take the dimensionless charge-to-mass ratio \(\xi=-10\). The initial
magnetic field is \(\vct{B}_0=(\cos87^\circ,0,\sin87^\circ)\), and the external
accelerations maintain zero center-of-mass acceleration with
\(\vct{a}_{\rm rel}=5\hat{\vct x}\). The gas and dust begin in the corresponding
equilibrium drift state. We denote the stopping time evaluated in this initial
equilibrium by \(t_{{\rm s},{\rm eq}}\) and use it to normalize time below. To seed the
instability, we perturb each gas and dust velocity component independently in every
cell, drawing the perturbations from a uniform distribution over \([-10^{-6},10^{-6}]\).

\begin{figure*}[p!]
\centering
\includegraphics[width=\textwidth]{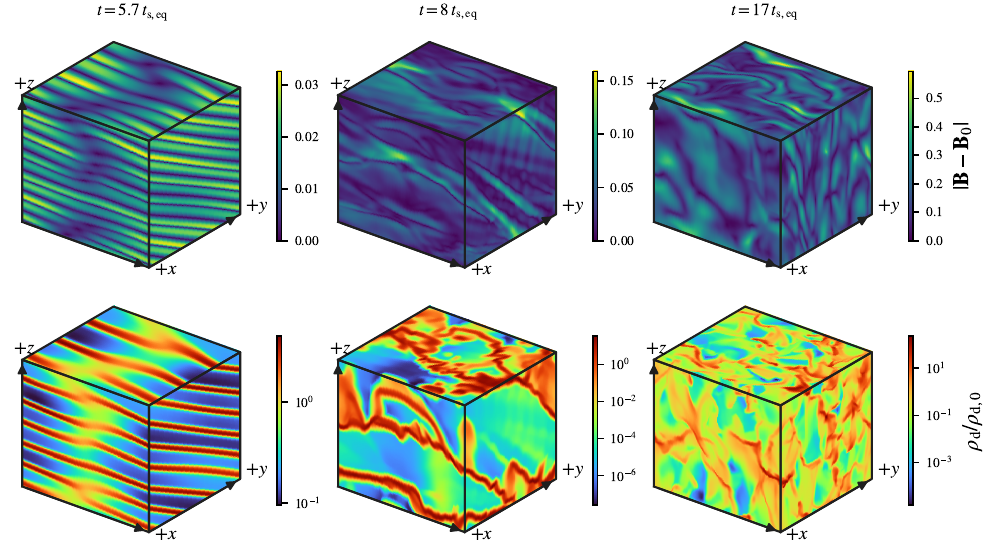}
\caption{Magnetic perturbation magnitude (top row) and normalized dust
density (bottom row) in the magnetized RDI during the linear, early
nonlinear, and saturated stages. Both quantities are shown on three
faces of the periodic box. The columns show \(t/t_{{\rm s},{\rm eq}}=5.7\), \(8\),
and \(17\), respectively.}
\label{fig:magnetized_rdi_stages}
\end{figure*}

\begin{figure*}[p!]
\centering
\includegraphics[width=\textwidth]{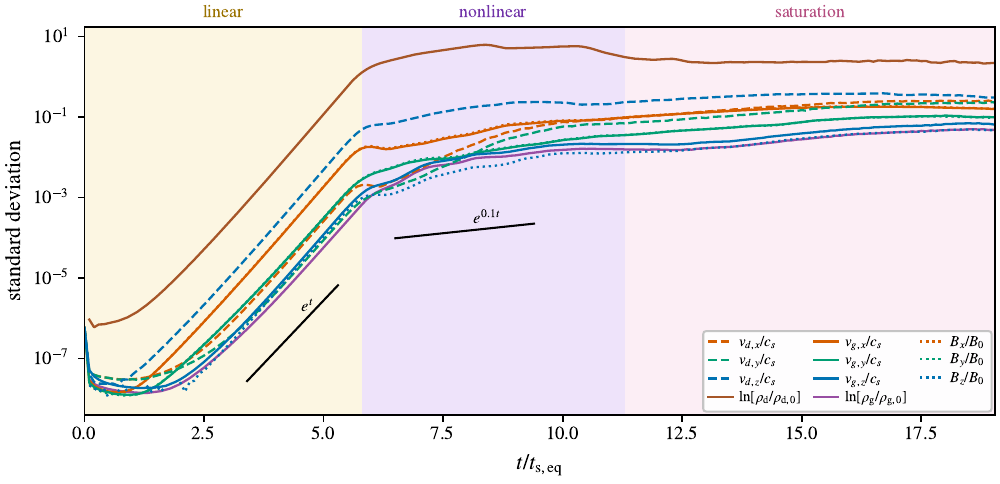}
\caption{Volume-weighted standard deviations of the logarithmic gas and dust densities,
velocity components, and magnetic field components in the magnetized RDI. The shaded
regions denote the linear, early nonlinear, and saturated stages, with boundaries at
\(t/t_{{\rm s},{\rm eq}}=5.8\) and \(11.3\). Velocities are normalized by \(c_s\), and
magnetic field components by \(B_0\). The black line segments show the \(e^t\) and
\(e^{0.1t}\) reference slopes used by \citet{Seligman2019}, with \(t\) in these
expressions measured in code units of \(L_{\rm box}/c_s\) and the slopes converted
to the time units used on the horizontal axis.}
\label{fig:magnetized_rdi_growth}
\end{figure*}

The evolution passes through rapid linear growth, slower early nonlinear growth, and a
statistically saturated state. We represent these stages with snapshots at
\(t/t_{{\rm s},{\rm eq}}=5.7\), \(8\), and \(17\), respectively.
Figure~\ref{fig:magnetized_rdi_stages} shows the magnetic and dust density structure at
these times. During the linear stage, the dust forms coherent, nearly planar sheets. In
the early nonlinear stage, the sheets narrow and corrugate, producing strong dust--gas
separation. The displayed range is approximately
\(10^{-7}\lesssim\rho_{\rm d}/\rho_{{\rm d},0}\lesssim50\). In saturation, the sheets
are distorted into a more irregular network and turbulent motions partly refill the
dust-poor regions. This progression is consistent with the morphology reported by
\citet{Seligman2019} and \citet{Moseley2023}.

Figure~\ref{fig:magnetized_rdi_growth} shows the volume-weighted standard deviations of
the logarithmic gas and dust densities, velocity components, and magnetic field
components. At the beginning of the calculation, several standard deviations briefly
decrease as fluctuations near the grid scale in the initial random perturbations are
rapidly damped. Once the unstable RDI modes dominate, the fluctuations grow rapidly
through the remainder of the linear stage, more slowly during the early nonlinear stage,
and then reach a statistically saturated state. The saturated gas remains subsonic and
weakly compressible, with \(\sigma(\ln\rho_{\rm g})\sim0.02\)--\(0.05\), whereas the
dust density is strongly nonlinear, with \(\sigma(\ln\rho_{\rm d})\sim2\)--\(3\). The
gas velocity and magnetic field fluctuations in each direction evolve similarly. In
saturation, the magnetic perturbations have an overall amplitude of order
\(\delta B/B_0\sim0.1\), consistent with the estimate of \citet{Seligman2019}.

\section{Discussion and summary}
\label{sec:summary_discussion}

\subsection{Discussion}

Tests of the source update confirm the analytic predictions for drag damping,
gyromotion errors, and terminal drift. In the nonlinear tests, Picard iteration
recovers fourth-order convergence for source integration with Gauss--Legendre
coefficients. The nonlinear solves require only a few iterations per source update
in these tests. Tests of dusty MHD waves, shocks, and a magnetized RDI further examine
the source update together with MHD and dust transport in spatially varying flows.

Because each dust species has only one velocity at a given position, the pressureless
Eulerian description cannot represent trajectory crossing or grains of the same species
within one cell at different phases of their gyromotion. If spurious dust-to-gas ratio
variations become significant in a particular application, the transport methods
developed for neutral dust by \citet{Verrier2025,Iwasaki2026} could in principle be
adapted to the present scheme. This is because our Strang splitting includes drag and
the Lorentz force in the source update, leaving the dust transport equations in the
same form as for neutral dust. Beyond the Eulerian multifluid method considered here,
the GIRK source integrator could in principle be adapted to a full one-fluid method by
formulating the local drag--Lorentz update in terms of the dust--gas differential
velocities used in the one-fluid formulation \citep{Hutchison2026}.

The gas and magnetic field are evolved under the ideal MHD approximation. The electric
field \(\vct E=-(\vg/c)\times\vct B\) therefore does not include corrections from the
current carried by charged dust.
In weakly ionized gas, charged grains can contribute to the electrical
conductivity and thereby affect the evolution of the magnetic field
\citep[e.g.,][]{WardleNg1999}. A generalized Ohm's law incorporating this feedback has
been used to simulate dust clumping \citep{VallucciGoy2025} and to analyze the linear
modes of a multifluid model retaining grain inertia \citep{Verrier2026ChargedMHD}.
Extending the present method to this regime would require coupling the charged dust
dynamics to a non-ideal MHD induction equation.

The method could also be coupled to a grain charging model. For example, an equilibrium
charging model based on the local plasma conditions and radiation field could provide
the state-dependent charge-to-mass ratios at the two RK stages. Other extensions
include dust production and grain size evolution through processes such as coagulation,
shattering, accretion, or sputtering, which would require additional source terms for
the dust populations. The dust fluids have no independent thermal energy equation,
and radiation forces on the dust are not calculated from the radiation field. A further
extension would calculate radiation forces and thermal exchange for the individual dust
species and include their effects on the gas and radiation field.

\subsection{Summary}

We have developed a two-stage general implicit Runge--Kutta (GIRK) source integrator
for multifluid charged dust coupled to an ideal MHD gas and implemented it in the
GPU-accelerated \textsc{quokka} code. Each dust species is represented by a separate
pressureless fluid. By treating aerodynamic drag and the Lorentz force implicitly,
we can use the CFL timestep even when the dust stopping time and gyromotion timescale
are much shorter. The main features of the method are as follows.

\begin{itemize}
\item Aerodynamic drag and the Lorentz force are advanced together in one implicit
source update, including their back-reaction on the gas. The gas and dust receive equal
and opposite momentum changes, conserving total gas--dust momentum. We combine this
source update with MHD and dust transport through Strang splitting, applying a source
half-step before and after each transport step.

\item The RK coefficients are selected by comparing the timestep with a characteristic
drag--Lorentz timescale determined by the stopping times and gyrofrequencies. For
timesteps below this timescale, we adopt the fourth-order Gauss--Legendre coefficients
when MHD is enabled, preserving the amplitude of drag-free gyromotion. For timesteps
at or above this timescale, the stiff coefficients damp unresolved relative motion
and approach the terminal drift set by the balance between drag, the Lorentz force,
and the different external forces acting on the gas and dust. For constant source
coefficients and external accelerations, the terminal drift error is
\(\mathcal O(\Delta t^{-2})\) in the large-timestep limit, where \(\Delta t\) is the
timestep.

\item The stopping time and charge-to-mass ratio can depend on the evolving state.
These source coefficients are evaluated separately at the two RK stage states.
We solve the resulting nonlinear stage equations by Picard iteration, reevaluating
the source coefficients from the updated stage states at each iteration.

\item We reduce the two-stage gas--dust momentum equations to a \(6\times6\) linear
system for the gas. The resulting gas stage solution is then substituted into the
equations for each dust species to obtain its momentum update. For \(N_{\rm d}\) dust
species, the complete linear solve costs \(\mathcal O(N_{\rm d})\), compared with
\(\mathcal O(N_{\rm d}^3)\) for a dense solve of the unreduced system. Each Picard
iteration retains this linear cost.

\item The gas energy update includes the change in gas kinetic energy and heating
from drag. Numerical integration can make the loss of total gas--dust kinetic energy
differ from the integrated physical drag heating. We separate this difference into
a drag contribution and a remaining gyrofrequency-dependent contribution. Including
the drag contribution restores total gas--dust energy conservation in the neutral
limit when all drag dissipation is deposited as gas heat. The Lorentz force exchanges
kinetic energy between gas and dust but does not produce physical heating, so our
default update does not deposit the gyrofrequency-dependent contribution as gas heat.
\end{itemize}

\begin{acknowledgments}
We thank Pinghui Huang for helpful discussions and Tianning Lyu for help with
debugging. ChatGPT was used to help draft and polish parts of the manuscript.
HL is supported by the National Key R\&D Program of China No. 2023YFB3002502,
the National Natural Science Foundation of China under No. 12373006 and 12533004,
and the China Manned Space Program with grant No. CMS-CSST-2025-A10.
CCH and MRK acknowledge support from the Australian Research Council through
Laureate Fellowship FL220100020. Development of \textsc{quokka} is supported by
computational resources provided by the Australian Government through the National
Computational Infrastructure and the Pawsey Supercomputing Centre under the National
Computational Merit Allocation Scheme.
\end{acknowledgments}

\appendix
\counterwithin*{equation}{section}

\section{Stability of the source update}
\label{app:stability}

For constant source coefficients, the stability of the drag-only GIRK update was
analyzed by \citet{TedeschiPrades2025}. With the Lorentz force included, the source
matrix \(\mathcal M\) defined by Equations~\eqref{eq:global_linear}
and~\eqref{eq:Mglobal} can have complex eigenvalues. We first show that these
eigenvalues lie in the closed left half-plane and then use this result to assess the
stability of the source update in the resolved and stiff branches.

For positive gas and dust densities, define the kinetic energy inner product of two
complex momentum states \(U\) and \(V\) by
\begin{equation}
  \langle U,V\rangle_E
  =
  \frac{\vct{p}_{{\rm g},U}^{*}\cdot\vct{p}_{{\rm g},V}}
       {\rho_{\rm g}}
  +
  \sum_i
  \frac{\vct{p}_{{\rm d},i,U}^{*}\cdot\vct{p}_{{\rm d},i,V}}
       {\rho_{{\rm d},i}} .
  \label{eq:app_energy_inner_product}
\end{equation}
Let \(V\) be an eigenvector of \(\mathcal M\) with eigenvalue \(\lambda\), and let
\(\vct w_i=\vct v_{{\rm d},i}-\vg\). The drag contribution to
\(\mathrm{Re}\,\langle V,\mathcal M V\rangle_E\) is
\(-\sum_i\rho_{{\rm d},i}\alpha_i|\vct w_i|^2\), while the Lorentz contribution has zero
real part because \(\Bcross^T=-\Bcross\). Therefore,
\begin{equation}
  \mathrm{Re}(\lambda)\|V\|_E^2
  =
  -\sum_i
  \rho_{{\rm d},i}\alpha_i|\vct w_i|^2
  \le 0 ,
  \label{eq:app_source_spectrum}
\end{equation}
where \(\|V\|_E^2=\langle V,V\rangle_E\). Thus every eigenvalue of \(\mathcal M\)
satisfies \(\mathrm{Re}(\lambda)\le0\).

The stability function \(R(z)\) for one \(\mathcal C\) half-step was introduced in
Section~\ref{subsec:relative_dynamics_diagnostics}. For an eigenmode of \(\mathcal M\)
with eigenvalue \(\lambda\), \(z=\Delta t_{\mathcal C}\lambda\), so
Equation~\eqref{eq:app_source_spectrum} gives \(\mathrm{Re}(z)\le0\). GL4, Midpoint, and
TP2025 are A-stable and therefore satisfy \(|R(z)|\le1\) throughout the closed left
half-plane. Thus none of the resolved coefficient sets amplifies an eigenmode of
\(\mathcal M\) when the source coefficients are constant.

The stiff coefficient set has the stability function
\(R_{\rm stiff}(z)=(1-z)/(1-2z+2z^2)\). It is not A-stable, but satisfies
\begin{equation}
  |R_{\rm stiff}(z)|\le1
  \quad\text{for}\quad
  \mathrm{Re}(z)\le0,\qquad |z|\ge\frac12.
  \label{eq:app_stiff_nonamplifying_region}
\end{equation}
To see this, write \(z=x+iy\). The difference between the squared moduli of the
denominator and numerator is \(|z|^2(4|z|^2-1)-8x|z|^2+8x^2-2x\), which is nonnegative
in this region. On the nonpositive real axis, write \(z=-a\) with \(a\ge0\). Then
\(R_{\rm stiff}(-a)=(1+a)/(1+2a+2a^2)\le1\), so these modes are stable for any timestep.

We use the separation of center-of-mass and relative motion introduced in
Section~\ref{subsec:relative_dynamics_diagnostics}. The center-of-mass velocity remains
unchanged during \(\mathcal C\), so its three independent velocity components correspond
to three zero eigenvalues of \(\mathcal M\). The remaining eigenvalues are those of
\(\mathcal M_{\rm rel}\). When \(\vct B=0\), the relative motion is governed by drag
alone in every direction. For nonzero \(\vct B\), motion parallel to the field is
also governed by drag alone. These modes have real, nonpositive eigenvalues and
are therefore stable under the stiff update.

For perpendicular motion, choose the \(z\)-axis along \(\vct B\). The relative
velocities satisfy
\(\dd\vct w_i/\dd t=\mtx T_i\vct w_i+\sum_k\epsilon_k\mtx T_k\vct w_k\). For the
perpendicular components, define \(u_{i,\pm}=w_{i,x}\pm i w_{i,y}\). Since
\((\mtx T_i\vct w_i)_x\pm i(\mtx T_i\vct w_i)_y =d_{i,\pm}u_{i,\pm}\), where
\(d_{i,\pm}=-\alpha_i\mp i\Omega_{{\rm L},i}\), these variables satisfy
\(\dd u_{i,\pm}/\dd t =d_{i,\pm}u_{i,\pm}+\sum_k\epsilon_k d_{k,\pm}u_{k,\pm}\).
Defining \(\mtx D_\pm=\mathrm{diag}(d_{1,\pm},\ldots,d_{N_{\rm d},\pm})\),
\(\vct\epsilon=(\epsilon_1,\ldots,\epsilon_{N_{\rm d}})^T\), and
\(\vct u_\pm=(u_{1,\pm},\ldots,u_{N_{\rm d},\pm})^T\), we obtain
\(\dd\vct u_\pm/\dd t=\mathcal M_{{\rm rel},\pm}\vct u_\pm\), where
\begin{equation}
  \mathcal M_{{\rm rel},\pm}
  =
  (\I+\vct 1\vct\epsilon^T)\mtx D_\pm,
  \label{eq:app_perpendicular_matrix}
\end{equation}
and \(\vct 1=(1,\ldots,1)^T\) is the vector of ones.

To obtain a lower bound on the eigenvalue magnitudes, we use a similarity transformation
with \(\mtx S=\mathrm{diag}(\sqrt{\epsilon_1},\ldots, \sqrt{\epsilon_{N_{\rm d}}})\). It
gives \(\mtx S\mathcal M_{{\rm rel},\pm}\mtx S^{-1} =\mtx P\mtx D_\pm\), where
\(\mtx P=\I+\vct s\vct s^T\) and
\(\vct s=(\sqrt{\epsilon_1},\ldots,\sqrt{\epsilon_{N_{\rm d}}})^T\). All eigenvalues of
the symmetric matrix \(\mtx P\) are at least unity, so
\(\|\mtx P\vct v\|_2\ge\|\vct v\|_2\) for any vector \(\vct v\). For an eigenvector
\(\vct y\) of \(\mtx P\mtx D_\pm\), this gives
\(|\lambda|\|\vct y\|_2 =\|\mtx P\mtx D_\pm\vct y\|_2 \ge\|\mtx D_\pm\vct y\|_2\). Since
\(\|\mtx D_\pm\vct y\|_2^2 =\sum_i(\alpha_i^2+\Omega_{{\rm L},i}^2)|y_i|^2\), every
perpendicular eigenmode satisfies
\begin{equation}
  |\lambda|\ge r_{\min},
  \qquad
  r_{\min}\equiv\min_i
  \sqrt{\alpha_i^2+\Omega_{{\rm L},i}^2}.
  \label{eq:app_relative_eigenvalue_bound}
\end{equation}

For constant source coefficients, Equation~\eqref{eq:tauDL} gives
\(\tau_{\rm DL}=r_{\min}^{-1}\). Whenever the stiff branch is selected, \(r_{\min}>0\)
and \(\Delta t\ge\tau_{\rm DL}\), so
\begin{equation}
  |z|=\frac{\Delta t}{2}|\lambda|
  \ge\frac{\Delta t}{2}r_{\min}\ge\frac12.
  \label{eq:app_stiff_branch_eigenvalue}
\end{equation}
Together with \(\mathrm{Re}(\lambda)\le0\), this places all perpendicular eigenmodes in
the stable region of Equation~\eqref{eq:app_stiff_nonamplifying_region}.

Thus the branch criterion ensures that no eigenmode of \(\mathcal M\) is amplified when
the source coefficients are constant. In the stiff branch, \(R_{\rm stiff}(0)=1\) and
\(R_{\rm stiff}(z)\to0\) as \(|z|\to\infty\), so the update preserves modes with zero
eigenvalues and damps those with nonzero eigenvalues in the large-timestep limit.

\section{RK energy residual of the source update}
\label{app:rk_energy}

Section~\ref{subsec:energy} defines the RK energy residual as the difference between the
energy correction required for discrete gas--dust energy conservation and the RK
integration of the physical drag heating. Here we derive this residual from the RK
momentum update and examine its leading-order behavior as \(\Delta t_{\mathcal C}\to0\)
in the resolved branch.

Using the kinetic energy inner product in Equation~\eqref{eq:app_energy_inner_product},
the kinetic energy of a momentum state \(U\) is \(\langle U,U\rangle_E/2\). For an
\(s\)-stage RK source update, let \(V_a=U^-+\Delta t_{\mathcal C}\sum_c a_{ac}K_c\),
\(K_a=F(V_a,E_{{\rm g},a})\), and \(U^+=U^-+\Delta t_{\mathcal C}\sum_a b_aK_a\).
The kinetic energy change over the source step satisfies
\begin{equation}
  \Delta E_{\rm kin}
  =
  \Delta t_{\mathcal C}\sum_a b_a\langle V_a,K_a\rangle_E
  -
  \frac{\Delta t_{\mathcal C}^2}{2}
  \sum_{a,c}m_{ac}\langle K_a,K_c\rangle_E,
  \label{eq:app_rk_energy_relation}
\end{equation}
where \(m_{ac}=b_a a_{ac}+b_c a_{ca}-b_a b_c\). Because \(\Bcross^T=-\Bcross\), the
Lorentz contribution to \(\langle V_a,K_a\rangle_E\) vanishes at each stage, while the
drag contribution is the negative of the physical drag heating rate. Using the
definition of \(\Delta E_{\rm cons}\) in Section~\ref{subsec:energy}, we have
\(\Delta E_{\rm cons}=-\Delta E_{\rm kin}\). Combining this with
Equation~\eqref{eq:app_rk_energy_relation} gives
\begin{equation}
  \Delta E_{\rm res}^{\rm RK}
  =
  \Delta E_{\rm cons}
  -
  \Delta E_{{\rm heat},{\rm RK}}^{\rm drag}
  =
  \frac{\Delta t_{\mathcal C}^2}{2}
  \sum_{a,c}m_{ac}\langle K_a,K_c\rangle_E .
  \label{eq:app_deltaE_residual}
\end{equation}

For the gas energy update, we separate this residual into drag and
gyrofrequency-dependent parts. At each stage, write \(K_a=K_a^{\rm drag}+K_a^{\rm L}\),
where the two terms are, respectively, the drag and Lorentz contributions to the
momentum stage rate. The drag residual is obtained by replacing \(K_a\) and \(K_c\) in
Equation~\eqref{eq:app_deltaE_residual} by \(K_a^{\rm drag}\) and \(K_c^{\rm drag}\).
The remaining contribution is
\(\Delta E_{\rm res}^{\Omega} =\Delta E_{\rm res}^{\rm RK}-\Delta E_{\rm res}^{\rm drag}\).
It contains the Lorentz and mixed drag--Lorentz contributions and vanishes when all
gyrofrequencies are zero.

For the two-stage tableau used here, \(a_{11}=\gamma_1\), \(a_{12}=\beta_1\),
\(a_{21}=\beta_2\), \(a_{22}=\gamma_2\), \(b_1=b\), and \(b_2=1-b\). The three
independent coefficients are \(m_{11}=2b\gamma_1-b^2\),
\(m_{22}=2(1-b)\gamma_2-(1-b)^2\), and \(m_{12}=b\beta_1+(1-b)\beta_2-b(1-b)\). For GL4
and Midpoint, all three coefficients vanish. For the TP2025 coefficients,
\((m_{11},m_{12},m_{22})=(1,-1/2,0)\), while the stiff coefficients give
\((m_{11},m_{12},m_{22})=(0,-1,1)\). Because all \(m_{ac}\) vanish for GL4 and Midpoint,
Equation~\eqref{eq:app_deltaE_residual} gives \(\Delta E_{\rm res}^{\rm RK}=0\)
identically, independent of the stage rates. The residual is generally nonzero for both
TP2025 and the stiff coefficients.

We next expand the residual as \(\Delta t_{\mathcal C}\to0\) in the resolved branch. For
a smooth source term, the RK stages satisfy
\(V_a-U^-=\mathcal O(\Delta t_{\mathcal C})\) and
\(K_a=K^-+\mathcal O(\Delta t_{\mathcal C})\), where
\(K^-=F(U^-,E_{\rm g}^-)\) is the momentum rate before the source update.
Equation~\eqref{eq:app_deltaE_residual} then gives
\begin{equation}
  \Delta E_{\rm res}^{\rm RK}
  =
  \frac{\Delta t_{\mathcal C}^2}{2}
  \left(\sum_{a,c}m_{ac}\right)
  \langle K^-,K^-\rangle_E
  +
  \mathcal O(\Delta t_{\mathcal C}^3).
  \label{eq:app_residual_small_step}
\end{equation}
The TP2025 coefficients satisfy \(\sum_a b_a=1\) and \(\sum_{a,c}b_a a_{ac}=1/2\).
Together with the definition of \(m_{ac}\), these relations give \(\sum_{a,c}m_{ac}=0\).
The \(\mathcal O(\Delta t_{\mathcal C}^2)\) term in
Equation~\eqref{eq:app_residual_small_step} therefore vanishes, and the TP2025 energy
residual begins at \(\mathcal O(\Delta t_{\mathcal C}^3)\).

For a general nonlinear source term, TP2025 is second-order accurate, so the truncation
error over one \(\mathcal C\) step scales as \(\mathcal O(\Delta t_{\mathcal C}^3)\).
Equation~\eqref{eq:app_residual_small_step} shows that the RK energy residual has the
same scaling and therefore does not introduce a lower-order term in
\(\Delta t_{\mathcal C}\). Because the gas energy update in Section~\ref{subsec:energy}
weights the drag and gyrofrequency-dependent residuals separately, we also require their
individual scalings. The drag residual has the same quadratic form as
Equation~\eqref{eq:app_deltaE_residual}, and the same expansion gives
\(\Delta E_{\rm res}^{\rm drag}=\mathcal O(\Delta t_{\mathcal C}^3)\). Since both
\(\Delta E_{\rm res}^{\rm RK}\) and \(\Delta E_{\rm res}^{\rm drag}\) are
\(\mathcal O(\Delta t_{\mathcal C}^3)\), their difference
\(\Delta E_{\rm res}^{\Omega}=\Delta E_{\rm res}^{\rm RK}-\Delta E_{\rm res}^{\rm drag}\)
is also \(\mathcal O(\Delta t_{\mathcal C}^3)\). Applying
Equation~\eqref{eq:app_rk_energy_relation} to the gas momentum alone likewise shows that
the difference between the gas kinetic energy change \(\Delta E_{{\rm g},{\rm kin}}\)
and the RK integration of gas mechanical work is \(\mathcal O(\Delta t_{\mathcal C}^3)\)
for TP2025 and vanishes identically for GL4 and Midpoint.

\section{Matrix formulas for the reduced implicit solve}
\label{app:algebra}

Section~\ref{subsec:reduction} reduces the full two-stage system by solving the dust
stage rates independently for each species in terms of the gas stage rates, leading to
Equation~\eqref{eq:dust_affine} and the \(6\times6\) gas system in
Equation~\eqref{eq:gas6}. Here we give the explicit matrix formulas for the dust system
of each species and for the gas system.

Let \(\mtx{\Pi}=\hat{\vct b}\hat{\vct b}^{T}\) and
\(\mtx{Q}=\I-\mtx{\Pi}\) be the orthogonal projectors onto the subspaces parallel and
perpendicular to the magnetic field, respectively. The identities
\(\Bcross^2=-\mtx{Q}\), \(\Bcross\mtx{\Pi}=\mtx{\Pi}\Bcross=\mtx{0}\),
\(\mtx{Q}\Bcross=\Bcross\mtx{Q}=\Bcross\),
\(\mtx{Q}\mtx{\Pi}=\mtx{\Pi}\mtx{Q}=\mtx{0}\), and
\(\mtx{Q}^2=\mtx{Q}\), \(\mtx{\Pi}^2=\mtx{\Pi}\) show that sums and products of
\(\mtx{Q}\), \(\Bcross\), and \(\mtx{\Pi}\) retain the same form. We write
\begin{equation}
  \mathcal{K}(a_\perp,b,a_\parallel)
  \equiv
  a_\perp\mtx{Q}
  +
  b\Bcross
  +
  a_\parallel\mtx{\Pi}.
  \label{eq:app_K_definition}
\end{equation}
For a vector parallel to the magnetic field, \(\mathcal{K}\) multiplies the vector by
\(a_\parallel\). For a vector \(\vct{x}_\perp\) perpendicular to the field,
\(\mathcal{K}\vct{x}_\perp
=a_\perp\vct{x}_\perp+b\Bcross\vct{x}_\perp\).

Matrices of this form commute. Their product and the inverse needed below are
\begin{align}
  \mathcal{K}(a_\perp,b,a_\parallel)
  \mathcal{K}(d_\perp,e,d_\parallel)
  &=
  \mathcal{K}\left(
    a_\perp d_\perp-be,
    a_\perp e+bd_\perp,
    a_\parallel d_\parallel
  \right),
  \notag\\
  \mathcal{K}(a_\perp,b,a_\parallel)^{-1}
  &=
  \mathcal{K}\left(
    \frac{a_\perp}{a_\perp^2+b^2},
    -\frac{b}{a_\perp^2+b^2},
    \frac{1}{a_\parallel}
  \right),
  \label{eq:app_K_algebra}
\end{align}
where the inverse exists when \(a_\perp^2+b^2\ne0\) and \(a_\parallel\ne0\).
In this notation, \(\I=\mathcal{K}(1,0,1)\) and
\(\mtx{T}_{a,i}=
\mathcal{K}(-\alpha_{a,i},\Omega_{{\rm L},a,i},-\alpha_{a,i})\).

Define \(\mtx{L}_{a,i}=\Delta t_{\mathcal C}\mtx{T}_{a,i}\). Let
\(\mathcal{S}_i\) denote the \(6\times6\) matrix on the left-hand side of
Equation~\eqref{eq:dust6}. Its four \(3\times3\) blocks have the form
\(\mathcal{K}(a_\perp,b,a_\parallel)\) and commute, so the inverse of
\(\mathcal{S}_i\) can be written directly in terms of the four blocks:
\begin{equation}
  \mathcal{S}_i^{-1}
  =
  \begin{pmatrix}
    (\I-\gamma_2\mtx{L}_{2,i})\mtx{\Delta}_i^{-1}
    &
    \beta_1\mtx{L}_{1,i}\mtx{\Delta}_i^{-1}\\
    \beta_2\mtx{L}_{2,i}\mtx{\Delta}_i^{-1}
    &
    (\I-\gamma_1\mtx{L}_{1,i})\mtx{\Delta}_i^{-1}
  \end{pmatrix},
  \label{eq:app_species_inverse}
\end{equation}
where
\(\mtx{\Delta}_i=(\I-\gamma_1\mtx{L}_{1,i})
(\I-\gamma_2\mtx{L}_{2,i})
-\beta_1\beta_2\mtx{L}_{1,i}\mtx{L}_{2,i}\).
The matrix \(\mtx{\Delta}_i\) also has the form
\(\mathcal{K}(a_\perp,b,a_\parallel)\), so its inverse is obtained from
Equation~\eqref{eq:app_K_algebra}. The matrices in Equation~\eqref{eq:dust_affine} are
then
\begin{equation}
  \begin{pmatrix}
    \mtx{P}_{1,i}\\
    \mtx{P}_{2,i}
  \end{pmatrix}
  =
  \mathcal{S}_i^{-1}
  \begin{pmatrix}
    \mtx{T}_{1,i}\\
    \mtx{T}_{2,i}
  \end{pmatrix},
  \label{eq:app_dust_P}
\end{equation}
and
\begin{equation}
  \begin{pmatrix}
    \mtx{X}_{1,i} & \mtx{Y}_{1,i}\\
    \mtx{X}_{2,i} & \mtx{Y}_{2,i}
  \end{pmatrix}
  =
  -\epsilon_i\mathcal{S}_i^{-1}
  \begin{pmatrix}
    \gamma_1\mtx{L}_{1,i} & \beta_1\mtx{L}_{1,i}\\
    \beta_2\mtx{L}_{2,i} & \gamma_2\mtx{L}_{2,i}
  \end{pmatrix}.
  \label{eq:app_dust_XY}
\end{equation}

The four \(3\times3\) blocks in Equation~\eqref{eq:gas6} are formed from \(\I\) and
sums of \(\mtx{X}_{a,i}\) and \(\mtx{Y}_{a,i}\). Because all dust species in a cell
use the same magnetic field direction, these blocks also have the form
\(\mathcal{K}(a_\perp,b,a_\parallel)\) and commute. Define
\(\mtx{\Lambda}_{11}=\I+\sum_i\mtx{X}_{1,i}\),
\(\mtx{\Lambda}_{12}=\sum_i\mtx{Y}_{1,i}\),
\(\mtx{\Lambda}_{21}=\sum_i\mtx{X}_{2,i}\), and
\(\mtx{\Lambda}_{22}=\I+\sum_i\mtx{Y}_{2,i}\), together with
\(\vct{r}_a=\sum_i\mtx{P}_{a,i}\qrel_i^-\). The solution of
Equation~\eqref{eq:gas6} is then
\begin{equation}
\begin{aligned}
  \mtx{\Delta}_{\rm g}
  &=
  \mtx{\Lambda}_{11}\mtx{\Lambda}_{22}
  -
  \mtx{\Lambda}_{12}\mtx{\Lambda}_{21},\\
  \vct{k}_{1,{\rm g}}
  &=
  -\mtx{\Delta}_{\rm g}^{-1}
  \left(
    \mtx{\Lambda}_{22}\vct{r}_1
    -
    \mtx{\Lambda}_{12}\vct{r}_2
  \right),\\
  \vct{k}_{2,{\rm g}}
  &=
  -\mtx{\Delta}_{\rm g}^{-1}
  \left(
    \mtx{\Lambda}_{11}\vct{r}_2
    -
    \mtx{\Lambda}_{21}\vct{r}_1
  \right).
\end{aligned}
\label{eq:app_gas_solution}
\end{equation}
The matrix \(\mtx{\Delta}_{\rm g}\) also has the form
\(\mathcal{K}(a_\perp,b,a_\parallel)\), so its inverse is evaluated using
Equation~\eqref{eq:app_K_algebra}. Once the gas stage rates are known, the dust stage
rates follow from Equation~\eqref{eq:dust_affine}. All matrix operations above have
fixed dimension, so increasing \(N_{\rm d}\) only adds the same matrix calculations for
each additional dust species and additional terms to the sums used to construct the
\(6\times6\) gas system. The total cost therefore scales as \(\mathcal O(N_{\rm d})\),
as described in Section~\ref{subsec:reduction}.

\bibliographystyle{aasjournalv7}
\bibliography{charged_dust_girk}

@article{BenitezLlambay2019,
  author = {Ben{\'i}tez-Llambay, Pablo and Krapp, Leonardo and Pessah, Martin E.},
  title = {{Asymptotically Stable Numerical Method for Multispecies Momentum Transfer: Gas and Multifluid Dust Test Suite and Implementation in FARGO3D}},
  journal = {ApJS},
  year = {2019},
  volume = {241},
  number = {2},
  pages = {25},
  doi = {10.3847/1538-4365/ab0a0e}
}

@article{Birnstiel2024,
  author = {Birnstiel, Tilman},
  title = {{Dust Growth and Evolution in Protoplanetary Disks}},
  journal = {ARA\&A},
  year = {2024},
  volume = {62},
  number = {1},
  pages = {157--202},
  doi = {10.1146/annurev-astro-071221-052705}
}

@article{Chapman2006,
  author = {Chapman, J. F. and Wardle, M.},
  title = {Dust grain dynamics in {C}-type shock waves in molecular clouds},
  journal = {MNRAS},
  year = {2006},
  volume = {371},
  pages = {513--529},
  doi = {10.1111/j.1365-2966.2006.10592.x}
}

@article{Commercon2023,
  author = {Commer\c{c}on, B. and Lebreuilly, U. and Price, D. J. and
            Lovascio, F. and Laibe, G. and Hennebelle, P.},
  title = {Dynamics of dust grains in turbulent molecular clouds.
           Conditions for decoupling and limits of different numerical implementations},
  journal = {A\&A},
  year = {2023},
  volume = {671},
  pages = {A128},
  doi = {10.1051/0004-6361/202245141}
}

@book{Draine2011,
  author = {Draine, B. T.},
  title = {{Physics of the Interstellar and Intergalactic Medium}},
  publisher = {Princeton University Press},
  address = {Princeton, NJ},
  year = {2011}
}

@article{Draine1980,
  author = {Draine, B. T.},
  title = {Interstellar shock waves with magnetic precursors},
  journal = {ApJ},
  year = {1980},
  volume = {241},
  pages = {1021--1038},
  doi = {10.1086/158416}
}

@incollection{Draine2004,
  author = {Draine, B. T.},
  title = {Astrophysics of Dust in Cold Clouds},
  booktitle = {The Cold Universe},
  editor = {Pfenniger, D. and Revaz, Y.},
  series = {Saas-Fee Advanced Courses},
  volume = {32},
  publisher = {Springer},
  address = {Berlin, Heidelberg},
  year = {2004},
  pages = {213--304},
  doi = {10.1007/3-540-31636-1_3},
  eprint = {astro-ph/0304488},
  archivePrefix = {arXiv}
}

@article{DraineSutin1987,
  author = {Draine, B. T. and Sutin, Brian},
  title = {{Collisional charging of interstellar grains}},
  journal = {ApJ},
  year = {1987},
  volume = {320},
  pages = {803--817},
  doi = {10.1086/165596}
}

@article{Dubois2024,
  author = {Dubois, Y. and Rodr\'{i}guez Montero, F. and Guerra, C. and
            Trebitsch, M. and Han, S. and Beckmann, R. and Yi, S. K. and
            Lewis, J. and Jang, J. K.},
  title = {Galaxies with grains: unraveling dust evolution and extinction
           curves with hydrodynamical simulations},
  journal = {A\&A},
  year = {2024},
  volume = {687},
  pages = {A240},
  doi = {10.1051/0004-6361/202449784}
}

@article{Guillet2007,
  author = {Guillet, V. and Pineau des For{\^e}ts, G. and Jones, A. P.},
  title = {{Shocks in dense clouds. I. Dust dynamics}},
  journal = {A\&A},
  year = {2007},
  volume = {476},
  number = {1},
  pages = {263--277},
  doi = {10.1051/0004-6361:20078094}
}

@article{HopkinsLee2016,
  author = {Hopkins, Philip F. and Lee, Hyunseok},
  title = {{The fundamentally different dynamics of dust and gas in molecular clouds}},
  journal = {MNRAS},
  year = {2016},
  volume = {456},
  number = {4},
  pages = {4174--4190},
  doi = {10.1093/mnras/stv2745}
}

@article{HuangBai2022,
  author = {Huang, P. and Bai, X.-N.},
  title = {A Multifluid Dust Module in {Athena++}: Algorithms and Numerical Tests},
  journal = {ApJS},
  year = {2022},
  volume = {262},
  pages = {11},
  doi = {10.3847/1538-4365/ac76cb}
}

@article{Krapp2024,
  author = {Krapp, Leonardo and Garrido-Deutelmoser, Juan and Ben{\'i}tez-Llambay, Pablo and Kratter, Kaitlin M.},
  title = {{A Fast Second-order Solver for Stiff Multifluid Dust and Gas Hydrodynamics}},
  journal = {ApJS},
  year = {2024},
  volume = {271},
  number = {1},
  pages = {7},
  doi = {10.3847/1538-4365/ad14f9}
}

@article{LaibePrice2012a,
  author = {Laibe, Guillaume and Price, Daniel J.},
  title = {{Dusty gas with smoothed particle hydrodynamics -- I. Algorithm and test suite}},
  journal = {MNRAS},
  year = {2012},
  volume = {420},
  number = {3},
  pages = {2345--2364},
  doi = {10.1111/j.1365-2966.2011.20202.x}
}

@article{LaibePrice2012b,
  author = {Laibe, Guillaume and Price, Daniel J.},
  title = {{Dusty gas with smoothed particle hydrodynamics -- II. Implicit timestepping and astrophysical drag regimes}},
  journal = {MNRAS},
  year = {2012},
  volume = {420},
  number = {3},
  pages = {2365--2376},
  doi = {10.1111/j.1365-2966.2011.20201.x}
}

@article{LaibePrice2014a,
  author = {Laibe, Guillaume and Price, Daniel J.},
  title = {{Dusty gas with one fluid}},
  journal = {MNRAS},
  year = {2014},
  volume = {440},
  number = {3},
  pages = {2136--2146},
  doi = {10.1093/mnras/stu355}
}

@article{LaibePrice2014b,
  author = {Laibe, Guillaume and Price, Daniel J.},
  title = {{Dusty gas with one fluid in smoothed particle hydrodynamics}},
  journal = {MNRAS},
  year = {2014},
  volume = {440},
  number = {3},
  pages = {2147--2163},
  doi = {10.1093/mnras/stu359}
}

@article{LaibePrice2014c,
  author = {Laibe, Guillaume and Price, Daniel J.},
  title = {{Dust and gas mixtures with multiple grain species -- a one-fluid approach}},
  journal = {MNRAS},
  year = {2014},
  volume = {444},
  number = {2},
  pages = {1940--1956},
  doi = {10.1093/mnras/stu1367}
}

@article{LeeHopkinsSquire2017,
  author = {Lee, H. and Hopkins, P. F. and Squire, J.},
  title = {The dynamics of charged dust in magnetized molecular clouds},
  journal = {MNRAS},
  year = {2017},
  volume = {469},
  pages = {3532--3540},
  doi = {10.1093/mnras/stx1097}
}

@article{McKinnon2018,
  author = {McKinnon, R. and Vogelsberger, M. and Torrey, P. and
            Marinacci, F. and Kannan, R.},
  title = {Simulating galactic dust grain evolution on a moving mesh},
  journal = {MNRAS},
  year = {2018},
  volume = {478},
  pages = {2851--2886},
  doi = {10.1093/mnras/sty1248}
}

@article{Mentiplay2020,
  author = {Mentiplay, D. and Price, D. J. and Pinte, C. and Laibe, G.},
  title = {A smoothed particle hydrodynamics algorithm for multigrain
           dust with separate sets of particles},
  journal = {MNRAS},
  year = {2020},
  volume = {499},
  pages = {3806--3818},
  doi = {10.1093/mnras/staa3171}
}

@article{Moseley2023,
  author = {Moseley, Eric R and Teyssier, R and Draine, B T},
  title = {{Dust dynamics in {RAMSES} -- I. Methods and turbulent acceleration}},
  journal = {MNRAS},
  year = {2023},
  volume = {518},
  number = {2},
  pages = {2825--2844},
  doi = {10.1093/mnras/stac3231},
  eprint = {2204.07681},
  archivePrefix = {arXiv}
}

@article{Moseley2025,
  author = {Moseley, Eric R and Teyssier, R},
  title = {{Dust dynamics in {RAMSES} -- II. Equilibrium drift velocity distributions of charged dust grains}},
  journal = {MNRAS},
  year = {2025},
  volume = {542},
  number = {2},
  pages = {1011--1032},
  doi = {10.1093/mnras/staf1260}
}

@article{PriceLaibe2015,
  author = {Price, Daniel J. and Laibe, Guillaume},
  title = {{A fast and explicit algorithm for simulating the dynamics of small dust grains with smoothed particle hydrodynamics}},
  journal = {MNRAS},
  year = {2015},
  volume = {451},
  number = {1},
  pages = {813--826},
  doi = {10.1093/mnras/stv996}
}

@article{TedeschiPrades2025,
  author = {Tedeschi-Prades, G. and Birnstiel, T. and Dolag, K. and
            Ercolano, B. and Hutchison, M.},
  title = {General Implicit {Runge--Kutta} Integrators for Multifluid
           Gas--Dust Aerodynamic Drag},
  journal = {ApJ},
  year = {2025},
  volume = {994},
  number = {2},
  pages = {267},
  doi = {10.3847/1538-4357/ae1ccc},
  eprint = {2511.07562},
  archivePrefix = {arXiv}
}

@article{Tricco2017,
  author = {Tricco, Terrence S. and Price, Daniel J. and Laibe, Guillaume},
  title = {{Is the dust-to-gas ratio constant in molecular clouds?}},
  journal = {MNRAS},
  year = {2017},
  volume = {471},
  number = {1},
  pages = {L52--L56},
  doi = {10.1093/mnrasl/slx096}
}

@article{Verrier2025,
  author = {Verrier, G. and Lebreuilly, U. and Hennebelle, P.},
  title = {{A new multifluid method for dusty astrophysical flows. Application to turbulent protostellar collapses}},
  journal = {A\&A},
  year = {2025},
  volume = {701},
  pages = {A174},
  doi = {10.1051/0004-6361/202554662}
}

@article{Wardle1998,
  author = {Wardle, M.},
  title = {Dust grains and the structure of steady {C}-type
           magnetohydrodynamic shock waves in molecular clouds},
  journal = {MNRAS},
  year = {1998},
  volume = {298},
  pages = {507--524},
  doi = {10.1046/j.1365-8711.1998.01641.x}
}

@article{WardleNg1999,
  author = {Wardle, Mark and Ng, Cindy},
  title = {{The conductivity of dense molecular gas}},
  journal = {MNRAS},
  year = {1999},
  volume = {303},
  number = {2},
  pages = {239--246},
  doi = {10.1046/j.1365-8711.1999.02211.x}
}

@article{Weidenschilling1977,
  author = {Weidenschilling, S. J.},
  title = {{Aerodynamics of solid bodies in the solar nebula}},
  journal = {MNRAS},
  year = {1977},
  volume = {180},
  number = {2},
  pages = {57--70},
  doi = {10.1093/mnras/180.2.57}
}

@article{WeingartnerDraine2001,
  author = {Weingartner, Joseph C. and Draine, B. T.},
  title = {{Photoelectric Emission from Interstellar Dust: Grain Charging and Gas Heating}},
  journal = {ApJS},
  year = {2001},
  volume = {134},
  number = {2},
  pages = {263--281},
  doi = {10.1086/320852}
}

@article{YoudinGoodman2005,
  author = {Youdin, Andrew N. and Goodman, Jeremy},
  title = {{Streaming Instabilities in Protoplanetary Disks}},
  journal = {ApJ},
  year = {2005},
  volume = {620},
  number = {1},
  pages = {459--469},
  doi = {10.1086/426895}
}

@article{Cuzzi2008,
  author = {Cuzzi, Jeffrey N. and Hogan, Robert C. and Shariff, Karim},
  title = {{Toward Planetesimals: Dense Chondrule Clumps in the Protoplanetary Nebula}},
  journal = {ApJ},
  year = {2008},
  volume = {687},
  number = {2},
  pages = {1432--1447},
  doi = {10.1086/591239}
}

@article{Johansen2007,
  author = {Johansen, Anders and Oishi, Jeffrey S. and Mac Low, Mordecai-Mark and Klahr, Hubert and Henning, Thomas and Youdin, Andrew},
  title = {{Rapid planetesimal formation in turbulent circumstellar disks}},
  journal = {Nature},
  year = {2007},
  volume = {448},
  number = {7157},
  pages = {1022--1025},
  doi = {10.1038/nature06086}
}

@article{Epstein1924,
  author = {Epstein, Paul S.},
  title = {{On the Resistance Experienced by Spheres in their Motion through Gases}},
  journal = {Physical Review},
  year = {1924},
  volume = {23},
  number = {6},
  pages = {710--733},
  doi = {10.1103/PhysRev.23.710}
}

@article{Baines1965,
  author = {Baines, M. J. and Williams, I. P. and Asebiomo, A. S.},
  title = {{Resistance to the Motion of a Small Sphere Moving Through a Gas}},
  journal = {MNRAS},
  year = {1965},
  volume = {130},
  number = {1},
  pages = {63--74},
  doi = {10.1093/mnras/130.1.63}
}

@article{Kwok1975,
  author = {Kwok, S.},
  title = {Radiation pressure on grains as a mechanism for mass loss
           in red giants},
  journal = {ApJ},
  year = {1975},
  volume = {198},
  pages = {583--591},
  doi = {10.1086/153637}
}

@article{WibkingKrumholz2022,
  author = {Wibking, Benjamin D and Krumholz, Mark R},
  title = {{{QUOKKA}: a code for two-moment AMR radiation hydrodynamics on GPUs}},
  journal = {MNRAS},
  year = {2022},
  volume = {512},
  number = {1},
  pages = {1430--1449},
  doi = {10.1093/mnras/stac439}
}

@article{Strang1968,
  author = {Strang, Gilbert},
  title = {{On the Construction and Comparison of Difference Schemes}},
  journal = {SIAM Journal on Numerical Analysis},
  year = {1968},
  volume = {5},
  number = {3},
  pages = {506--517},
  doi = {10.1137/0705041}
}

@article{Seligman2019,
  author = {Seligman, D. and Hopkins, P. F. and Squire, J.},
  title = {{Non-linear evolution of the resonant drag instability in magnetized gas}},
  journal = {MNRAS},
  year = {2019},
  volume = {485},
  pages = {3991--3998},
  doi = {10.1093/mnras/stz666}
}

@article{Sewanou2025,
  author = {Sewanou, L. and Laibe, G. and Commer{\c c}on, B.},
  title = {{A highly accurate drag solver for multi-fluid dust and gas hydrodynamics on GPUs}},
  journal = {A\&A},
  year = {2025},
  volume = {703},
  pages = {A59},
  doi = {10.1051/0004-6361/202556530}
}

@article{Iwasaki2026,
  author = {Iwasaki, K.},
  title = {{A Split-Flux Method for Dust--Gas Two-Fluid Simulations from Strong to Weak Coupling}},
  journal = {arXiv e-prints},
  year = {2026},
  eid = {arXiv:2607.24418},
  eprint = {2607.24418},
  archivePrefix = {arXiv},
  doi = {10.48550/arXiv.2607.24418}
}

@article{VallucciGoy2025,
  author = {Vallucci-Goy, V. and Hennebelle, P. and Lebreuilly, U. and Verrier, G.},
  title = {{Magnetic clumping of charged dust in the dense interstellar medium}},
  journal = {A\&A},
  year = {2025},
  volume = {704},
  pages = {A142},
  doi = {10.1051/0004-6361/202554503}
}

@article{Hutchison2026,
  author = {Hutchison, Mark and Laibe, Guillaume and Tedeschi-Prades, Giovanni and David-Cl{\'e}ris, Timoth{\'e}e and Barret, Alex and Lombart, Maxime and Price, Daniel J and Koepferl, Christine M},
  title = {{Full one-fluid dusty gas with multiple grain species in SPH}},
  journal = {MNRAS},
  year = {2026},
  volume = {550},
  number = {2},
  pages = {stag1106},
  doi = {10.1093/mnras/stag1106},
  eprint = {2606.10676},
  archivePrefix = {arXiv}
}

@article{Miniati2010,
  author = {Miniati, Francesco},
  title = {{A hybrid scheme for gas--dust systems stiffly coupled via viscous drag}},
  journal = {Journal of Computational Physics},
  year = {2010},
  volume = {229},
  number = {10},
  pages = {3916--3937},
  doi = {10.1016/j.jcp.2010.01.034},
  eprint = {1001.4794},
  archivePrefix = {arXiv}
}

@article{CiolekRoberge2002,
  author = {Ciolek, Glenn E. and Roberge, Wayne G.},
  title = {{Time-dependent, Multifluid, Magnetohydrodynamic Shock Waves with Grain Dynamics. I. Formulation and Numerical Tests}},
  journal = {ApJ},
  year = {2002},
  volume = {567},
  number = {2},
  pages = {947--961},
  doi = {10.1086/338591},
  eprint = {astro-ph/0112569},
  archivePrefix = {arXiv}
}

@article{HopkinsSquire2018,
  author = {Hopkins, Philip F and Squire, Jonathan},
  title = {{Ubiquitous instabilities of dust moving in magnetized gas}},
  journal = {MNRAS},
  year = {2018},
  volume = {479},
  number = {4},
  pages = {4681--4719},
  doi = {10.1093/mnras/sty1604},
  eprint = {1801.10166},
  archivePrefix = {arXiv}
}

@article{Lebreuilly2019,
  author = {Lebreuilly, U. and Commer\c{c}on, B. and Laibe, G.},
  title = {Small dust grain dynamics on adaptive mesh refinement grids.
           {I}. Methods},
  journal = {A\&A},
  year = {2019},
  volume = {626},
  pages = {A96},
  doi = {10.1051/0004-6361/201834147},
  eprint = {1905.01948},
  archivePrefix = {arXiv}
}

@article{DraineSalpeter1979Drag,
  author = {Draine, B. T. and Salpeter, E. E.},
  title = {On the physics of dust grains in hot gas},
  journal = {ApJ},
  year = {1979},
  volume = {231},
  pages = {77--94},
  doi = {10.1086/157165}
}

@article{Verrier2026ChargedMHD,
  author = {Verrier, Gabriel and Hennebelle, Patrick and
            Lebreuilly, Ugo and Vallucci-Goy, Valentin},
  title = {Magnetohydrodynamics of charged interstellar dust.
           Multifluid models and study of the linear modes},
  journal = {arXiv e-prints},
  year = {2026},
  eprint = {2608.30711},
  archivePrefix = {arXiv},
  primaryClass = {astro-ph.GA},
  note = {Accepted for publication in Astronomy \& Astrophysics},
  url = {https://arxiv.org/abs/2608.30711}
}

@article{BeitiaAntero2021,
  author = {Beitia-Antero, Leire and
            G{\'o}mez de Castro, Ana I. and Vallejo, Juan C.},
  title = {Formation of Dust Filaments in the Diffuse Envelopes
           of Molecular Clouds},
  journal = {ApJ},
  year = {2021},
  volume = {908},
  number = {1},
  pages = {112},
  doi = {10.3847/1538-4357/abcda1},
  eprint = {2008.13135},
  archivePrefix = {arXiv}
}

@article{HopkinsSquireSeligman2020,
  author = {Hopkins, Philip F. and Squire, Jonathan and
            Seligman, Darryl},
  title = {Simulating diverse instabilities of dust in magnetized gas},
  journal = {MNRAS},
  year = {2020},
  volume = {496},
  number = {2},
  pages = {2123--2154},
  doi = {10.1093/mnras/staa1046}
}

@article{He2024a,
  author = {He, Chong-Chong and Wibking, Benjamin D. and Krumholz, Mark R.},
  title = {{An asymptotically correct implicit--explicit time integration scheme for finite volume radiation-hydrodynamics}},
  journal = {MNRAS},
  year = {2024},
  volume = {531},
  number = {1},
  pages = {1228--1242},
  doi = {10.1093/mnras/stae1244},
  eprint = {2404.08247},
  archivePrefix = {arXiv}
}

@article{He2024b,
  author = {He, Chong-Chong and Wibking, Benjamin D. and Krumholz, Mark R.},
  title = {{A novel numerical method for mixed-frame multigroup radiation-hydrodynamics with GPU acceleration implemented in the QUOKKA code}},
  journal = {MNRAS},
  year = {2024},
  volume = {535},
  number = {4},
  pages = {3059--3076},
  doi = {10.1093/mnras/stae2580},
  eprint = {2407.18304},
  archivePrefix = {arXiv}
}

@article{Kriel2026,
  author = {Kriel, Neco and Cole-Kodikara, Elizabeth and Wibking, Benjamin and
            Krumholz, Mark R. and He, Chong-Chong and Sarkar, Anish and
            Li, Pak Shing},
  title = {{Magnetising the QUOKKA code with fast, second-order accurate,
            error-correcting schemes for magnetohydrodynamics on GPUs}},
  journal = {arXiv e-prints},
  year = {2026},
  eid = {arXiv:2609.10816},
  eprint = {2609.10816},
  archivePrefix = {arXiv},
  primaryClass = {astro-ph.IM}
}

@article{Parente2026,
  author = {Parente, Massimiliano and Narayanan, Desika and Torrey, Paul},
  title = {{Dust and Grain Size Evolution in Galaxy Simulations:
            What Matters and What Does Not}},
  journal = {The Open Journal of Astrophysics},
  year = {2026},
  volume = {9},
  doi = {10.33232/001c.166984},
  eprint = {2604.06314},
  archivePrefix = {arXiv},
  primaryClass = {astro-ph.GA}
}

@article{Squire2021,
  author = {Squire, Jonathan and Hopkins, Philip F. and Quataert, Eliot and
            Kempski, Philipp},
  title = {The impact of astrophysical dust grains on the confinement of cosmic rays},
  journal = {MNRAS},
  year = {2021},
  volume = {502},
  number = {2},
  pages = {2630--2644},
  doi = {10.1093/mnras/stab179},
  eprint = {2011.02497},
  archivePrefix = {arXiv}
}

\end{document}